\documentclass[a4paper,graphics,natbib,graphicx,psfig,amssymb,ccaption,layout]{article}
\usepackage{lscape, morefloats , graphicx, float}
\newcommand{\affil}[1]{$^{\rm #1}$}
\renewcommand{\baselinestretch}{.95}
\usepackage[authoryear]{natbib}
\bibpunct{(}{)}{;}{a}{}{,}
\date{} 

\title{\large\bf\flushleft Local stellar kinematics from RAVE data-
V. Kinematic investigation of the Galaxy with red clump stars}
\author{\parbox{\textwidth}{\flushleft
\vspace{-0.5cm}
{\it S. Karaali\affil{A, *}, S. Bilir\affil{A}, S. Ak\affil{A}, E. Yaz G\"ok\c ce\affil{A}, \"O. \"Onal\affil{A}, T. Ak\affil{A}}\\
\vspace{0.4cm}
{\small \affil{A}\,Istanbul University, Faculty of Sciences, Department of Astronomy and Space Sciences, 34119, Istanbul, Turkey}\\
{\small \affil{*}\,Email: karsa@istanbul.edu.tr}}}
\begin{document}
\twocolumn[
\begin{changemargin}{.8cm}{.5cm}
\begin{minipage}{.9\textwidth}
\vspace{-1cm}
\maketitle
\small{\bf Abstract:}
We investigated the space velocity components of 6610 red clump (RC) stars in 
terms of vertical distance, Galactocentric radial distance and Galactic 
longitude. Stellar velocity vectors are corrected for differential rotation 
of the Galaxy which is taken into account using photometric distances of RC 
stars. The space velocity components estimated for the sample stars above and 
below the Galactic plane are compatible only for the space velocity component 
in the direction to the Galactic rotation of the thin disc stars. The space 
velocity component in the direction to the Galactic rotation ($V_{lsr}$) shows 
a smooth variation relative to the mean Galactocentric radial distance ($R_m$), 
while it attains its maximum at the Galactic plane. The space velocity 
components in the Galactic centre ($U_{lsr}$) and in the vertical direction 
($W_{lsr}$) show almost flat distributions relative to $R_m$ with small changes 
in their trends at $R_m \sim 7.5$ kpc. $U_{lsr}$ values estimated for the RC 
stars in quadrant $180^{o}<l\leq270^{o}$ are larger than the ones in quadrants 
$0^{o}<l\leq90^{o}$ and $270^{o}<l\leq360^{o}$. The smooth distribution of the 
space velocity dispersions reveals that the thin and thick discs are 
kinematically continuous components of the Galaxy. Based on the $W_{lsr}$ 
space velocity components estimated in the quadrants $0^{o}<l\leq90^{o}$ and 
$270^{o}<l\leq360^{o}$, in the inward direction relative to the Sun, we showed 
that RC stars above the Galactic plane move towards the North Galactic Pole, 
whereas those below the Galactic plane move in the opposite direction. In the 
case of quadrant $180^{o}<l\leq270^{o}$, their behaviour is different, i.e. 
the RC stars above and below the Galactic plane move towards the Galactic plane. 
We stated that the Galactic long bar is the probable origin of many, but not 
all, of the detected features.

\medskip
\medskip
\end{minipage}
\end{changemargin}
]
\small

\section{Introduction}
Several studies based on different techniques and data revealed the non-steady 
state and asymmetrical structure of our Galaxy. The Milky Way Galaxy is still 
evolving under the effects of internal and external forces. After the discovery 
of the accretion of the Sagittarius dwarf galaxy \citep*{Ibata94}, researchers 
drew their attention to the Galactic streams. Sagittarius stream 
\citep{Majewski03} is associated with the Sagittarius dwarf galaxy. However, 
there are Galactic streams whose origins are not yet known. Some of them are 
tidal debris and some of them originate from the accretion. Galactic warp and 
dynamical interaction of the thick disc with the Galactic long bar can be 
associated with  some of the Galactic streams \citep[][and the references 
therein]{Williams13}. The presence of some of the streams are revealed by their 
large-scale stellar over-densities. Monoceros stream \citep{Newberg02, Yanny03}
outward from the Sun and Hercules thick disc cloud \citep*{Larsen96, Parker03, 
Parker04, Larsen08} inward from the Sun are examples for these over-density 
structures. Helmi stream \citep{Helmi99} and the recent Aquarius stream 
\citep{Williams11} are also two notable streams.  

Star count analysis is one of the procedures used to reveal the complex 
structure of the Galaxy. \citet{Bilir06} showed that the scaleheights and the 
scalelengths of the thin and thick discs are Galactic longitude dependent. 
In \citet{Ak07a,Ak07b}, the metallicities for relatively short vertical 
distances ($z<2.5$ kpc) show systematic fluctuations with Galactic longitude 
which was interpreted as the flare effect of the disc. A more comprehensive 
study was carried out by \citet{Bilir08} who showed that the thin and thick 
disc scaleheights as well as the axis ratio of the halo varies with Galactic 
longitude. The variation of these parameters were explained with the 
gravitational effect of the Galactic long bar. A similar work is carried 
out by \citet{Yaz10} in intermediate latitudes of the Galaxy where the 
variations of the thin and thick disc scaleheights were explained with the 
effect of the disc flare and disc long bar. 

The velocity distribution in the UV plane is also complex, i.e. it differs 
from a smooth Schwarzschild distribution. This has been proven for the solar 
neighbourhood \citep{Dehnen98} and the recent studies revealed that the same 
case holds also for the solar suburb \citep{Antoja12}. The complex structure 
is probably created by the Galactic long bar and spiral arms. Dissolving open 
clusters and perturbative effect of the disc by merger events can also be 
used for the explanation of the complexity \citep{Williams13}. \citet{Siebert11} 
showed that the radial velocities ($V_R$) estimated via the RAdial Velocity 
Experiment \citep[RAVE;][]{Steinmetz06} data are non-zero and also they have 
a small gradient, i.e. $dV_R/dR<-3$ kms$^{-1}$kpc$^{-1}$. A similar result 
based on RAVE red clump (RC) stars has been cited in \citet{Casetti11}. 
\citet{Siebert12} used  density-wave models to show that the radial streaming 
originates from the resonance effect of the spiral arms, and reproduced the 
gradient just mentioned. Probably, the most comprehensive study is that of 
\citet{Williams13} which is based on the stars from the internal release 
of RAVE data in  October 2011. Beyond a detailed error analysis, \citet{Williams13} 
confirmed the radial gradient $dV_{R}/dR<-3$ kms$^{-1}$kpc$^{-1}$ and they revealed 
the different behaviour of the vertical velocities, $V_Z$, of the RC stars in 
opposite regions of $R\sim8$ kpc in the ($R$, $Z$) plane. Also, \citet{Williams13} 
argued that the Hercules thick disc cloud \citep{Larsen96, Parker03, Larsen08} is an 
important phenomenon which causes the variation of the stellar velocities. 
\citet{Parker03} give the Galactic coordinates of the thick disc cloud as 
$20^{o}\leq l\leq 55^{o}$, $25^{o}\leq b\leq 45^{o}$, $305^{0}\leq l\leq 340^{o}$, 
and $-45^{o}\leq b\leq -25^{o}$. \citet{Larsen08} stated that the center of the 
overdensity region ranges from ($X$, $Y$, $Z$)=(6.5, -2.2, 1.5) to 
($X$, $Y$, $Z$)=(6.5, 0.3, 1.5) kpc, and that there is a clear excess of stars 
in quadrant $Q_1$ over quadrant $Q_4$ in the range $l=25^{o}-45^{o}$ and 
$b=30^{o}-40^{o}$.    

In this study, we intend to contribute to the discussions of the complexity 
of the Milky Way Galaxy by investigating the variation of the space velocity 
components of 6610 RC stars. The difference between the procedures in the 
literature and ours is the application of a series of constraints in our work, 
i.e. 1) we used the space velocity components instead of the cylindrical 
coordinates; 2) we applied corrections for the differential rotation to our 
velocities; 3) we investigated the variation of the velocities for three 
Galactic populations, thin and thick discs, halo and their combinations; 
4) we investigated the lag of the sample stars relative to the local standard 
of rest stars; and 5) we investigated the variation of the velocities in terms 
of $z_{min}$ and $z_{max}$ instead of current positions which covers the 
effect of long lived internal and external forces. The paper is organized as 
follows. The data are given in Section 2. Section 3 is devoted to the 
distribution of the velocity components for stars relative to several 
parameters: i) vertical distance, i.e. for stars above and below the Galactic 
plane separately, ii) Galactocentric distance, and iii) Galactic longitude. 
The distribution of the velocity dispersions for different velocities is 
also given in this section. Finally, a discussion of the results and a short 
conclusion is presented in Section 4.

\section{Data}
The data of 6781 RC stars are taken from \citet{Bilir12}. They used 
the RAVE Data Release 3 \citep[DR3,][]{Siebert11} survey and applied a series 
of constraints to 83 072 radial velocity measurements to identify 7985 RC stars 
among them. Also, they carried out the following evaluations to obtain the 
final sample of RC stars: The proper motions of 7846 stars were taken from 
RAVE DR3 while the 139 stars which were not available in this 
catalogue were provided from the PPMXL catalogue of \citet*{Roeser10}. 
Distances were obtained by combining the apparent $K_{S}$ magnitude of the star 
in question and the absolute magnitude $M_{K_{S}}=-1.54\pm0.04$ mag, adopted 
for all RC stars \citep{Groenewegen08}, while the $E(B-V)$ reddening was 
obtained iteratively by using published methodology \citep[cf.][and references 
therein]{Coskunoglu11}. The apparent $K_{S}$ magnitudes were de-reddened 
by means of the equations in \citet{Fiorucci03}. Distribution of the RC stars 
in Equatorial and Galactic coordinates is given in Fig. 1. 

\citet{Bilir12} combined the distances with the RAVE DR3 radial velocities and 
available proper motions, applying the algorithms and transformation matrices 
of \citet{Johnson87} to obtain the Galactic space velocity components $(U,V,W)$ 
of the sample stars. In the calculations, epoch $J2000$ was adopted as described 
in the International Celestial Reference System of the {\em Hipparcos} and 
{\em Tycho-2} catalogues \citep{ESA97}. The transformation matrices use the 
notation of right-handed system. Hence, $U$, $V$ and $W$ are the components of 
a velocity vector of a star with respect to the Sun, where $U$ is positive 
towards the Galactic center ($l=0^{o}$, $b=0^{o}$), $V$ is positive in the 
direction of the Galactic rotation ($l=90^{o}$, $b=0^{o}$) and $W$ is positive 
towards the North Galactic Pole ($b=90^{o}$).    

\citet{Bilir12} adopted the value of the rotation speed of the Sun as 222.5 
kms$^{-1}$. Correction for differential Galactic rotation is necessary for 
accurate determination of the $U$, $V$ and $W$ velocity components. The effect 
is proportional to the projection of the distance to the stars on to the 
Galactic plane, i.e. the $W$ velocity component is not affected by Galactic 
differential rotation \citep{Mihalas81}. They applied the procedure of 
\citet{Mihalas81} to the distribution of the sample stars and estimated the 
first-order Galactic differential rotation corrections for the $U$ and $V$ 
velocity components of the sample stars. The $U$, $V$ and $W$ velocities were 
reduced to local standard of rest (LSR) by adopting the solar LSR velocities 
in \citet{Coskunoglu11}, ($U_\odot$, $V_\odot$, $W_\odot$)=(8.83, 14.19, 6.57) 
kms$^{-1}$. We will use the symbols $U_{lsr}$, $V_{lsr}$ and $W_{lsr}$ for them, 
hereafter. The uncertainties of the space velocities $U_{err}$, $V_{err}$ and 
$W_{err}$ (Fig. 2) were computed by propagating the uncertainties of the proper 
motions, distances and radial velocities, again using an algorithm by 
\citet{Johnson87}. Then, the error for the total space motion of a star follows 
from the equation

\begin{eqnarray}
S_{err}^2=U_{err}^2+V_{err}^2+W_{err}^2.
\end{eqnarray}

\citet{Bilir12} removed the RC stars with total space velocity errors larger 
than the mean errors ($\langle S_{err}\rangle=39$ kms$^{-1}$) plus the standard deviation 
($\sigma=36$ kms$^{-1}$), i.e. $S_{err}>75$ kms$^{-1}$, thus the sample reduced 
to 6781 stars. Also in this study, 171 RC stars that are very close to Galactic 
plane ($-10^{o}\leq b\leq +10^{o}$) were excluded from the sample of \citet{Bilir12}. 
These stars are in the calibration fields, so their properties such as age may be 
different from the general Galactic population we intend to study here. Thus, 
the final sample used in this study reduced to 6610. The large errors originate 
from the proper motions. The proper motions of 706 stars with $S_{err}>75$ kms$^{-1}$ 
is $\mu_{tot_{err}}\leq10$ mas yr$^{-1}$ and those of 498 stars is 
$\mu_{tot_{err}}>10$ mas yr$^{-1}$. The distance histogram of the RC 
stars (Fig. 3) shows that those with large errors, $S_{err}>75$ kms$^{-1}$, 
locate at large distances. A proper motion error of 10 mas yr$^{-1}$ 
corresponds to 50 kms$^{-1}$ at 1 kpc, and correspondingly more if further 
away. Hence, omitting RC stars with $S_{err}>75$ kms$^{-1}$ removed the space 
velocity components with large errors. Also, distance and radial velocity 
errors may affect the space velocity components. However, in our study, they 
are small, i.e. distances are based on the absolute magnitude 
$M_{K_{s}}=-1.54\pm$0.04 mag \citep{Groenewegen08}, where the error is rather 
small, and RAVE group gives a median radial velocity error of 
1.2 kms$^{-1}$ \citep{Siebert11}. The distribution of the sample stars in the 
($X$, $Y$) and ($X$, $Z$) planes and their space velocity components ($U$, $V$) 
and ($W$, $V$) are plotted in Fig. 4 and Fig. 5, respectively. 
Both figures involve the sample stars  as well as the rejected ones. 
The RC stars with $S_{err}>75$ kms$^{-1}$ (blue colour) locate at the outermost 
region of Fig. 4, i.e. their $X$, $Y$ and $Z$ coordinates are larger than the 
sample stars (grey colour), while the RC stars close to the Galactic plane,
$|b|<10^{o}$, occupy the central part of the figure, as expected. The positions 
of the sample stars and the stars close to the Galactic plane in Fig. 5 are 
almost the same as in Fig. 4. However, the stars with  errors 
$S_{err}>75$ kms$^{-1}$ -and with large distances- are concentrated in the 
central part of the diagram giving the indication that their (relatively) 
large errors reduced their space velocity components to smaller values.

We used standard gravitational potentials described in the literature 
\citep{Miyamoto75, Hernquist90, Johnston95, Dinescu99} to estimate orbital 
elements of each of the sample stars. The orbital elements for a star used 
in our work are the mean of the corresponding orbital elements calculated 
over 15 orbital periods of that specific star. The orbital integration 
typically corresponds to 3 Gyr and is sufficient to evaluate the orbital 
elements of solar suburb stars \citep{Coskunoglu12, Bilir12, Duran13}. 

First, we performed the test-particle integration in a Milky Way potential 
which consists of a logarithmic halo to determine a possible orbit in the 
form below:
\begin{eqnarray}
  \Phi_{\rm halo}(r)=v_{0}^{2} \ln \left(1+\frac{r^2}{d^2}\right),
\end{eqnarray}
with $v_{0}=186$ kms$^{-1}$ and $d=12$ kpc. The disc is represented
by a Miyamoto-Nagai potential \citep{Miyamoto75}:
\begin{eqnarray}
  \Phi_{\rm disc}(R,z)=-\frac{G M_{\rm d}} { \sqrt{R^{2} + \left(
        a_d + \sqrt{z^{2}+b_d^{2}} \right)^{2}}},
\end{eqnarray}
with $M_{\rm d}=10^{11}~M_{\odot}$, $a_d=6.5$ kpc and $b_d=0.26$
kpc. Finally, the bulge is modeled as a Hernquist potential \citep{Hernquist90}
\begin{eqnarray}
  \Phi_{\rm bulge}(r)=-\frac{G M_{\rm b}} {r+c},
\end{eqnarray}
using $M_{\rm b}=3.4\times10^{10}~M_{\odot}$ and $c=0.7$ kpc. The
superposition of these components gives quite a good representation of
the Milky Way. The circular speed at the solar radius is 222.5 kms$^{-1}$. 
$P_{LSR}=2.18\times10^8$ years is the orbital period of the
LSR and $V_c=222.5$ kms$^{-1}$ denotes the circular rotational
velocity at the solar Galactocentric distance, $R_0=8$ kpc.  

For our kinematic analysis, we are interested in the mean radial Galactocentric 
distance ($R_{m}$) as a function of the stellar population and the orbital 
shape. \citet{Williams11} have analyzed the radial orbital eccentricities of 
RAVE sample of thick-disc stars, to test thick-disc formation models. Here, 
we consider the vertical orbital eccentricity, $e_v$ for population analysis. 
$R_m$ is defined as the arithmetic mean of the final 
perigalactic ($R_p$) and apogalactic ($R_a$) distances, and $z_{max}$ and 
$z_{min}$ are the final maximum and minimum values of the $z$ coordinates, 
respectively, to the Galactic plane, where $e_v$ is defined as follows:
\begin{eqnarray}
e_v=\frac{(|z_{max}|+|z_{min}|)}{R_m}
\end{eqnarray}
where $R_m=(R_a+R_p)/2$ \citep{Pauli05}. 

\section{Distribution of the Space Velocity Components}
\subsection{Distribution of the space velocity components above 
and below the Galactic plane}
We adopted the vertical orbital eccentricities and the procedure in 
\citet{Bilir12} and separated all the sample into three populations, 
thin disc ($e_v\leq0.12$), thick disc ($0.12<e_v\leq 0.25$) and halo 
($e_v>0.25$). We carried out the same evaluation for the RC stars above 
and below the Galactic plane. The Galactic latitudes of these sub-samples 
are $b>10^{o}$ and $b<-10^{o}$, respectively, due to the restriction 
explained in Section 2. The space velocity components and their corresponding 
dispersions for three categories are given in Table 1. The number of stars 
for the thin disc are in majority, while those for the halo are in minority, 
as expected for a sample of stars in the solar suburb. The data confirm 
another expectation of us, i.e. the numerical values for a specific velocity 
component are different for different populations. One can see in Table 1 
that there is a symmetrical distribution relative to the Galactic plane for 
only two parameters, i.e. the space velocity component $V_{lsr}$ and its 
total dispersion $\sigma_{tot}$. The number of stars below the Galactic plane 
is larger than the ones above, due to the observational strategy of RAVE, 
3801 and 2809 stars respectively. Hence, the errors of the space velocity 
components for the stars with $b<-10^{o}$ are less than the corresponding 
ones with $b>10^{o}$.    
 
\subsection{Distribution of the space velocity components relative to the 
Galactocentric radial distance in different $z_{min}$ and $z_{max}$ intervals}

We estimated the space velocity components of the sample stars relative to 
$R_m$ in different $z_{min}$ and $z_{max}$ intervals. The ranges of these 
parameters are $4<R_m\leq11$, $-2.5\leq z_{min}\leq0$ and 
$0<z_{max}\leq2.5$ kpc. The results are given in Table 2. The distributions 
of the space velocity components are given in Fig. 6. In the following, we 
discuss the trends in each space velocity component. 

\subsubsection{$U_{lsr}$}
The  variation of the space velocity component $U_{lsr}$ is given at the top panel 
of Fig. 6. When we consider the errors, the general trend is a flat distribution. 
However, there are different distributions in some $z_{min}$/$ z_{max}$ intervals, 
such as $-1<z_{min}\leq-0.5$ and $-0.5<z_{min}\leq 0$ kpc where a small 
increase in $U_{lsr}$ can be detected at $\sim$7.5 kpc. Whereas, the numerical 
value of $U_{lsr}$ in the interval $1.5<z_{max}\leq 2$ kpc at distance $R_m \sim 7.5$ 
kpc give the indication of a change in the trend, i.e. the decreasing velocity 
component in the distance interval $5<R_{m}<7.5$ kpc flattens at larger $R_m$ 
distances. A different figure is related to the extreme intervals, 
$-2.5\leq z_{min}\leq-2$ and $2<z_{max}\leq2.5$ kpc, where $U_{lsr}$ is an increasing 
function of $R_m$. However, the number of stars in these intervals are small. 

\subsubsection{$V_{lsr}$}
The middle panel in Fig. 6 shows that there is a smooth variation of the velocity 
component $V_{lsr}$ with respect to $R_m$ for all $z_{min}$ and $z_{max}$ intervals. 
Also, there is a slight indication (at least for the smallest values in each interval) 
that $V_{lsr}$ values for the RC stars for $b<-10^{o}$ increases with decreasing distance 
to the Galactic plane, while they increase gradually with increasing distance to the 
Galactic plane for $b>10^{o}$. This argument has been confirmed by the mean of $V_{lsr}$ 
velocity components in $z_{min}$ and $z_{max}$ intervals. Fig. 7 shows that there is 
almost a symmetrical distribution of the mean $V_{lsr}$ velocity components with respect 
to $z_{min}$ and $z_{max}$ for all RC stars. That is the $V_{lsr}$ velocity component 
values are highest in the Galactic plane and they become relatively lower with higher 
Galactic latitudes. This is what we expect from the Jeans equations, $V_{lsr}$ decreases 
as the asymmetrical drift increases and asymmetric drift increases with velocity 
dispersion \citep[cf.][]{Binney98}. The symmetric shape of the curve in this 
figure is also a general property of stellar Galactic orbits, i.e. 
$z_{min} \approx -z_{max}$, due to symmetry of the Galactic potential 
relative to the Galactic plane.   

\subsubsection{$W_{lsr}$}
The distribution of the space velocity component $W_{lsr}$ is given in the lower 
panel of Fig. 6. The distribution is rather flat in all $z_{min}$ and $z_{max}$ 
intervals, except in the intervals $-2.5\leq z_{min}\leq -2$ and $1.5<z_{max}\leq 2$ kpc 
where in the first interval there is a concave shape with a minimum at $\sim$7.5 kpc 
and where there is an extended peak covering the Galactocentric distances larger than 
$R_m \sim 7.5$ kpc following the flat distribution in shorter distances, $R_{m} \leq 7.5$
kpc. While $W_{lsr}$ increases with $R_m$ distance monotonously in the interval 
$2<z_{max}\leq 2.5$ kpc. We omit the first bin in this interval which contains 
only five stars. 

\subsection{Distribution of the space velocity components relative to the
Galactic longitude in different $z_{min}$ and $z_{max}$ intervals}
In this section, we discuss the distribution of the space velocity components 
relative to the Galactic longitude. However, we stress that we can not discuss 
motions in all four quadrants on equal footing, as RAVE observed only stars in 
the southern celestial hemisphere. We evaluated the mean of the space velocity 
components $U_{lsr}$, $V_{lsr}$ and $W_{lsr}$, for four quadrants, 
$Q_1(0^{o}\leq l\leq90^{o})$, $Q_2(90^{o}<l\leq180^{o})$, $Q_3(180^{o}<l\leq270^{o})$
and $Q_4(270^{o}<l\leq360^{o})$ of the Galaxy (Table 3) and plotted them in Fig. 8. 
We discuss the most striking features of the space velocity components in the following.

\subsubsection{$U_{lsr}$}
Distribution of the $U_{lsr}$ space velocity components relative to the 
quadrants are plotted at the top panel in Fig. 8. Two features can be detected 
in $z_{min}$ and $z_{max}$ intervals: i) There are systematic differences between the 
space velocity components in the quadrants $Q_1$, $Q_3$ and $Q_4$, $U_{lsr}$ being larger 
in $Q_3$ than ones in $Q_1$ and $Q_4$ in all $z_{min}$ and $z_{max}$ intervals, except 
$2<z_{max}\leq 2.5$ kpc. ii) The $U_{lsr}$ space velocity component corresponding 
the data in $Q_1$ decreases monotonously with increasing distance to the 
Galactic plane in the ranges $-1.5<z_{min}\leq0$ and $0<z_{max}\leq 1.5$ kpc 
and it increases at relatively extreme distances. 

\subsubsection{$V_{lsr}$}
The middle panel of Fig. 8 shows that, for $z_{min}$ intervals, the $V_{lsr}$ 
space velocity component for a given $Q_i$ ($i$=1, 3, 4) increases with increasing 
$z_{min}$, i.e. a result stated in Section 3.2.1. That is the $V_{lsr}$ space velocity 
components attain their larger values at lower Galactic latitudes. Any numerical value 
of $V_{lsr}$ which does not obey to this argument is related to the number of stars 
used for its evaluation and consequently its error. As an example, we give the $V_{lsr}$ 
values and the number of stars for the intervals $-1.5<z_{min}\leq-1$, $-2<z_{min}\leq-1.5$, 
and $-2.5\leq z_{min}\leq-2$ kpc, for the quadrant $Q_1$, i.e. $V_{lsr}=-27.7\pm20.08$ kms$^{-1}$, 
$N=246$; $V_{lsr}=-14.54\pm22.62$ kms$^{-1}$, $N=103$; and $V_{lsr}=-23.32\pm27.12$ kms$^{-1}$, 
$N=57$, respectively. The numerical value of $V_{lsr}$ for the first $z_{min}$ interval obeys 
the argument just cited, whereas those for the other two $z_{min}$ intervals with higher 
errors and less number of stars do not. For $z_{max}$ intervals, the same case holds only 
for the $Q_3$ and $Q_4$, while the behaviour of the $V_{lsr}$ space velocity 
component for $Q_1$ is in opposite sense. We can not consider the $V_{lsr}$ space velocity 
component for $Q_2$ due to absence of stars for this quadrant.   

\subsubsection{$W_{lsr}$}
Distribution of the $W_{lsr}$ space velocity components relative to the quadrants 
are plotted in the lower panel of Fig. 8. The behaviours of $W_{lsr}$ for stars in 
two quadrants are rather interesting. In $Q_1$, $W_{lsr}$ is positive in three 
$z_{max}$ intervals, $0<z_{max}\leq 0.5$, $0.5<z_{max}\leq 1$, 
and $1<z_{max}\leq 1.5$ kpc, while it is negative in all five $z_{min}$ intervals. 
That is, the RC stars above the Galactic plane in $Q_1$ move towards the direction 
of North Galactic Pole, whereas those below the Galactic plane in the same quadrant 
move in the opposite direction. The number of stars in the intervals $1.5<z_{max}\leq 2$ 
and $2<z_{max}\leq 2.5$ kpc which do not obey the argument just stated are only N=16 and 
6, respectively. However, the case is reverse for the RC stars in quadrant $Q_3$, 
i.e. $W_{lsr}$ is negative in the same $z_{max}$ intervals, $0<z_{max}\leq 0.5$, 
$0.5<z_{max}\leq 1$, and $1<z_{max}\leq 1.5$ kpc, while it is positive in three 
$z_{min}$ intervals: $-2<z_{min}\leq-1.5$, $-1.5<z_{min}\leq-1$, $-1<z_{min}\leq-0.5$ 
kpc. That is, the RC stars above and below the Galactic plane in the $z_{max}$ and 
$z_{min}$ intervals cited move towards the Galactic plane.      

The errors of the $U_{lsr}$ and $W_{lsr}$ are larger than the absolute values of the 
corresponding space velocities. Hence, the behaviours of these velocity components 
are probably more complicated than we detected. RAVE observed only stars in the 
southern celestial hemisphere. Hence, the number of stars with $z_{max}$ distances 
are relatively smaller than the corresponding ones with $z_{min}$ ones which causes 
larger errors. 

\subsection{Distribution of the space velocity dispersions}
We estimated the dispersions of the space velocity components for the 
sample stars as a function of $R_m$ in different $z_{min}$ and $z_{max}$ 
intervals. The ranges of $R_m$, $z_{min}$ and $z_{max}$ are $4\leq R_m\leq11$, 
${-2.5}\leq z_{min}\leq0$ and $0< z_{max}\leq2.5$ kpc, respectively. The results 
are given in Table 2 and Fig. 9. The RC stars at extreme Galactocentric distances, 
$4\leq R_m\leq5$ and $10<R_m\leq11$ kpc, are small in number. 
The number of stars are also relatively small for ${-2.5}\leq z_{min}\leq{-2}$ 
and $2<z_{max}\leq2.5$ kpc intervals, and their errors are large. In our 
discussions below, we will give less weight to the bins corresponding to small 
number of stars and relatively large errors.

The distribution of the $\sigma_U$ dispersion is almost flat in
${-0.5}<z_{min}\leq0$ and $0<z_{max}\leq0.5$ kpc intervals, while there is 
a positive (but small) gradient in other $z_{min}$ and $z_{max}$ 
intervals, i.e. $\sigma_U$ increases with increasing $R_m$. That is, the 
distribution of $\sigma_U$ is different between the ranges close to the 
Galactic plane and the further ones. 

The distribution of the $\sigma_V$ dispersion is also flat in 
${-0.5}<z_{min}\leq0$ and $0<z_{max}\leq0.5$ kpc intervals. However, one can detect 
a small gradient in some of the other $z_{min}$ and $z_{max}$ intervals, such 
as ${-1.5}< z_{min}\leq{-1}$ and ${0.5}< z_{max}\leq{1}$kpc.

The trend of the distribution of the dispersion $\sigma_W$ is almost the 
same as $\sigma_U$ and $\sigma_V$ in the intervals  ${-0.5}<z_{min}\leq0$ 
and $0<z_{max}\leq0.5$ kpc, while there is a gradient in the distribution 
of $\sigma_W$  in the other $z_{min}$ and $z_{max}$ intervals. Additionally, 
this gradient is in opposite sense cited for $\sigma_U$ velocity dispersion , 
i.e. $\sigma_W$ decreases with increasing $R_m$ distance.

The general aspect of the distribution of each space velocity dispersion, $\sigma_U$, 
$\sigma_V$, $\sigma_W$ and $\sigma_{tot}$, in Fig. 9 gives the indication of a 
parabolic function with its vertex at the Galactic plane: the space velocity dispersions 
are small in two intervals, ${-0.5}<z_{min}\leq0$ and $0<z_{max}\leq0.5$ kpc, 
but gradually, they become larger in the $z_{min}$ and $z_{max}$ intervals correspond to a 
larger distances from the Galactic plane. The small gradients in the $z_{min}$ 
and $z_{max}$ intervals encouraged us to estimate the mean space velocity dispersion in 
these intervals and plotted them versus corresponding mean $z_{min}$ and $z_{max}$ values. 
Fig. 10 shows that a Gaussian distribution fits to each of the space velocity 
dispersion, i.e. $\sigma_U$, $\sigma_V$, $\sigma_W$, $\sigma_{tot}$. Also, 
the smooth distribution of the velocity dispersions reveals that the thin 
and thick discs are kinematically continuous components of the Galaxy. 
Actually, the majority of our sample consists of the thin and thick discs 
RC stars, and there is a smooth transition between the small (thin disc) 
and relatively large (thick disc) space velocity dispersion in Fig. 10.     
As stated in Section 3.2.2, the symmetric shapes of the curves in 
this figure is also a general property of stellar Galactic orbits, i.e. 
$z_{min} \approx -z_{max}$, and the symmetric of the Galactic potential relative 
to the Galactic plane.   

\section {Summary and Discussion}
We used the space velocity components and their dispersions of 6610 RC stars and 
investigated their distribution relative to vertical distance, Galactocentric radial 
distance and Galactic longitude. The total error of the space velocity components is 
restricted with $S_{err}\leq75$ kms$^{-1}$, and space velocity components are corrected 
with differential rotation, to obtain reliable data. Also, we investigated the variation 
of the space velocity components in terms of $z_{min}$ and $z_{max}$ distances instead of 
$z$ which covers the effect of long lived internal and external forces.

In our distance calculations, we adopted a single value of absolute value, 
$M_{K_{S}}=-1.54$ mag \citep{Groenewegen08}. It has been cited in the 
literature \citep[cf.][]{Williams13} that the use of a single value does not 
compromise the results considerably. The proper motions of all RC stars are taken 
from RAVE DR3, except those of 139 stars which were not available in this 
catalogue and which were provided from the PPMXL, as cited in Section 2. 
Different  proper motion catalogues such as SPM4 and UCAC3 could be used as 
well. However, different proper motions change only the predicted values of 
the space velocity components but not the trends of their distributions. 
Additionally, the difference between the values of a given space velocity 
component predicted by two different proper motion catalogue is not large 
\citep[cf.][]{Williams13}. We omitted the RC stars with total error 
$S_{err}>75$ kms$^{-1}$ in their space velocities where most of these errors 
originate from the proper motions. The proper motions of 706 stars with 
$S_{err}>75$ kms$^{-1}$ is $\mu_{tot_{err}}\leq10$ mas and those of 498 
stars is $\mu_{tot_{err}}>10$ mas. The distances are based on a single 
absolute magnitude with an error of $\pm0.04$ mag, and RAVE group gives 
a median radial velocity error of 1.2 kms$^{-1}$ \citep{Siebert11}. 
Hence, the effect of the distance errors and radial velocity should be 
much smaller than the one for proper motion. 

The space velocity components and their dispersions for different populations, 
i.e. thin and thick discs and halo, are different as expected. The space 
velocity components for RC stars above ($b>10^{o}$) and below ($b<-10^{o}$) 
the Galactic plane are compatible only for $V_{lsr}$ of the thin disc.

The $V_{lsr}$ space velocity component is vertical distance ($z_{min}$ and $z_{max}$) 
and $R_m$ dependent. There is a smooth variation relative to $R_m$. The mean of the 
$V_{lsr}$ space velocity components for 10 $z_{min}$ and $z_{max}$ intervals could be 
fitted to a Gaussian distribution with a minimum point at the Galactic plane, a result 
expected from the Jeans equations, i.e. $V_{lsr}$ decreases as the asymmetric drift 
increases and asymmetric drift increases with space velocity dispersion \citep{Binney98}.

The general trend of the space velocity component $U_{lsr}$ is a flat distribution in terms 
of $R_m$ for most of the $z_{min}$ and $z_{max}$ intervals. However, there are some intervals 
where the trend changes at $\sim$7.5 kpc. Also, the distribution in the extreme intervals, 
${-2.5}\leq z_{min}\leq{-2}$ and $2<z_{max}\leq2.5$ kpc is an increasing function. 
However, the number of stars in these intervals are relatively small corresponding to 
larger errors.

The distribution of the space velocity component $W_{lsr}$ is also flat, again with 
some exceptions, i.e. there is a concave shape with a minimum at $R_m \sim 7.5$ kpc 
in the interval $-2.5\leq z_{min}\leq-2$ kpc, and an extended peak covering distances 
larger than $R_m \sim 7.5$ kpc, following the flat distribution in shorter distances. 
The distribution of $W_{lsr}$ is an increasing function in the extreme interval, 
$2<z_{max}\leq2.5$, similar to $U_{lsr}$.  

The behaviours of the space velocity components are different in four quadrants (Fig. 11). 
The upper panel of Fig. 11 shows that the space velocity components $U_{lsr}$ estimated 
for the RC stars in the quadrant $Q_3$ ($180^{o}<l\leq270^{o}$) are larger than the 
ones estimated for the RC stars in quadrants $Q_1$ ($0^{o}\leq l\leq90^{o}$) and $Q_4$ 
($270^{o}<l\leq360^{o}$) for all $z_{min}$ and $z_{max}$ intervals, except the 
interval $2<z_{max}\leq2.5$ kpc which is not valid for $Q_4$. We will see in the 
following that $z_{min}$ and $z_{max}$ intervals which involves small number of stars, 
such as in the present case where the number of stars is only $N=15$, show exceptions 
due to corresponding large errors. Different $U_{lsr}$ space velocity components 
indicates different lag for the RC stars in different quadrants. The $U_{lsr}$ 
velocity components for quadrants $Q_1$ and $Q_4$ attain their maximum values 
at small distances to the Galactic plane, while its distribution for the quadrant
$Q_3$ gives the indication of a double peak distribution. In our analysis, 
we give less weight to the (relatively) extreme $z_{min}$ and $z_{max}$ intervals 
due to larger errors, as mentioned above in this paragraph.

The space velocity component $V_{lsr}$ shows almost a similar distribution 
in terms of $z_{min}$ and $z_{max}$ distances in the quadrants $Q_3$ and $Q_4$ 
(middle panel), i.e. $V_{lsr}$ attains its (relative algebraic) large numerical 
values in $-0.5< z_{min}\leq0$ and $0<z_{max}\leq1$ kpc intervals, then it 
decreases gradually with larger distances from the Galactic plane. That is, 
the lag for the space velocity component in the direction to the Galactic rotation 
increases from the Galactic plane to larger vertical distances. However, the 
same case does not hold for $V_{lsr}$ estimated for the RC stars in the quadrant 
$Q_1$, i.e. the distribution of $V_{lsr}$ shows two minima and three maxima 
in the interval formed by combining five $z_{min}$ and five $z_{max}$ 
intervals. Especially, $V_{lsr}$ is a monotonous increasing function in the $z_{max}$ 
intervals, starting from a minimum point in $0<z_{max}\leq0.5$ kpc interval
which also contradicts with the maximum $V_{lsr}$ space velocity components 
estimated in the quadrants $Q_3$ and $Q_4$.

In the case of the space velocity component $W_{lsr}$, the similarity exists 
for the RC stars in the quadrants $Q_1$ and $Q_4$, as explained in the following. 
The numerical values of $W_{lsr}$ for the quadrants $Q_1$ and $Q_4$ are negative 
for all $z_{min}$ intervals, while they are positive for $Q_1$ in three $z_{max}$ 
intervals, $0<z_{max}\leq0.5$, $0.5<z_{max}\leq1$, and $1<z_{max}\leq1.5$ kpc. 
However, the number of stars in the other two $z_{max}$ intervals where $W_{lsr}$ 
is not positive, are only 16, and 6. $W_{lsr}$ is positive for $Q_4$ for four 
$z_{max}$ intervals, and its numerical value in the fifth $z_{max}$ interval, 
$0.5<z_{max}\leq1$ kpc, is only -0.73 kms$^{-1}$. The overall picture for the 
space velocity component in the quadrants $Q_1$ and $Q_4$ (in the inward 
direction relative to the Sun) is that RC stars above the Galactic plane move 
in the direction to the North Galactic Pole, whereas those below the Galactic 
plane move in the opposite direction, they show a rarefaction. The distribution 
of the space velocity component $W_{lsr}$ in the quadrant $Q_3$ is different 
than the ones in the quadrants $Q_1$ and $Q_4$, i.e. $W_{lsr}$ is positive 
(or close to zero) in four $z_{min}$ intervals and negative in four $z_{max}$ 
intervals. That is, in the case of quadrant $Q_3$ (in the outward direction), 
RC stars above and below the Galactic plane move towards the Galactic plane, 
they show a compression. We could not consider the quadrant $Q_2$ due to absence 
of stars in the $z_{max}$ intervals and only 94 stars in the $z_{min}$ intervals. 
Also, we stress (again) that we can not discuss motions in three quadrants on 
equal footing, as RAVE observed only stars in the southern celestial hemisphere. 

Our results confirm the complex structure of our Galaxy. It is non-homogeneous, 
non-steady state, and asymmetric. The rarefaction and the compression of the RC 
stars in different quadrants  were detected also in \citet{Williams13}. However, their 
procedure is different. In \citet{Williams13}, two opposite features were based 
on the Galactocentric radial distance, i.e. inward of $R_o=8.5$ kpc stars show 
a rarefaction, while outward of $R_o$ kpc they show a compression. In \citet{Bilir08}, 
the Galactic longitude dependence of the scaleheights of the thin and thick discs were 
explained with the gravitational effect of the Galactic long bar 
\citep[see also,][]{Cabrera-Lavers07}. Hence, the features detected in 
our study probably originate from the Galactic long bar, rather than any 
other event such as an accretion. The vertical distances, $z_{min}$ and 
$z_{max}$, used in this study correspond to a long lived time scale. 
Hence, any event related to the gradients of the space velocity components 
should be also long lived. Also this argument favors the Galactic long bar.  
      
However, we do not argue that all the features mentioned above can be explained 
with only one event. Differences between the lags for the $U_{lsr}$ space 
velocity components of the stars at the same vertical distance but in 
different quadrants, such as $Q_1$, $Q_3$ and $Q_4$, need alternative events. 

{\bf Conclusion:} The distributions of the space velocity components $U_{lsr}$, 
$V_{lsr}$, and $W_{lsr}$, relative to the vertical distance $z_{max}/z_{min}$, 
Galactocentric distance $R$, and Galactic longitude $l$ show that the 
Milky Way Galaxy has a non-homogeneous, non-steady state and asymmetric 
structure. The Galactic long bar is the probable origin of many, but not all, 
of the features detected.  

\section{Acknowledgments}
We are grateful to the anonymous reviewer who improved our paper by his/her 
comments and suggestions.This work has been supported in part by the 
Scientific and Technological Research Council (T\"UB\.ITAK) 112T120.   
Funding for RAVE has been provided by: the Australian Astronomical Observatory; 
the Leibniz-Institut fuer Astrophysik Potsdam (AIP); the Australian National 
University; the Australian Research Council; the French National Research 
Agency; the German Research Foundation; the European Research Council 
(ERC-StG 240271 Galactica); the Istituto Nazionale di Astrofisica at Padova; 
The Johns Hopkins University; the National Science Foundation of the USA 
(AST-0908326); the W. M. Keck foundation; the Macquarie University; the 
Netherlands Research School for Astronomy; the Natural Sciences and 
Engineering Research Council of Canada; the Slovenian Research Agency; the 
Swiss National Science Foundation; the Science \& Technology Facilities 
Council of the UK; Opticon; Strasbourg Observatory; and the Universities of 
Groningen, Heidelberg and Sydney. The RAVE web site is at\\ 
http://www.rave-survey.org 

This publication makes use of data products from the Two Micron All
Sky Survey, which is a joint project of the University of
Massachusetts and the Infrared Processing and Analysis
Center/California Institute of Technology, funded by the National
Aeronautics and Space Administration and the National Science
Foundation.  This research has made use of the SIMBAD, NASA's
Astrophysics Data System Bibliographic Services and the NASA/IPAC
ExtraGalactic Database (NED) which is operated by the Jet Propulsion
Laboratory, California Institute of Technology, under contract with
the National Aeronautics and Space Administration.

\pagebreak[3]

\begin{figure}
\begin{center}
\includegraphics[scale=0.43, angle=0]{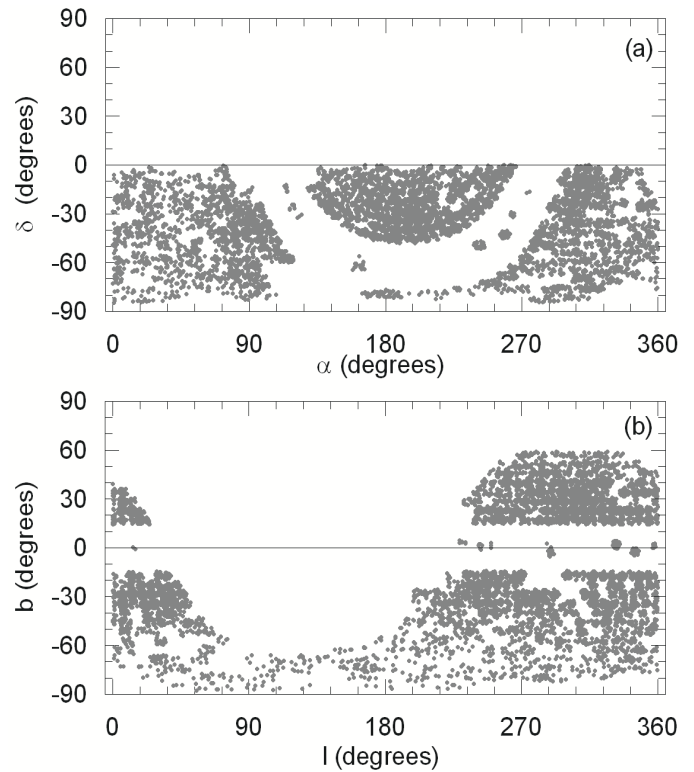}
\caption[] {Distribution of RAVE DR3 RC stars in the Equatorial (a)
and Galactic (b) coordinates.}
\end{center}
\end{figure} 

\begin{figure}
\begin{center}
\includegraphics[scale=0.43, angle=0]{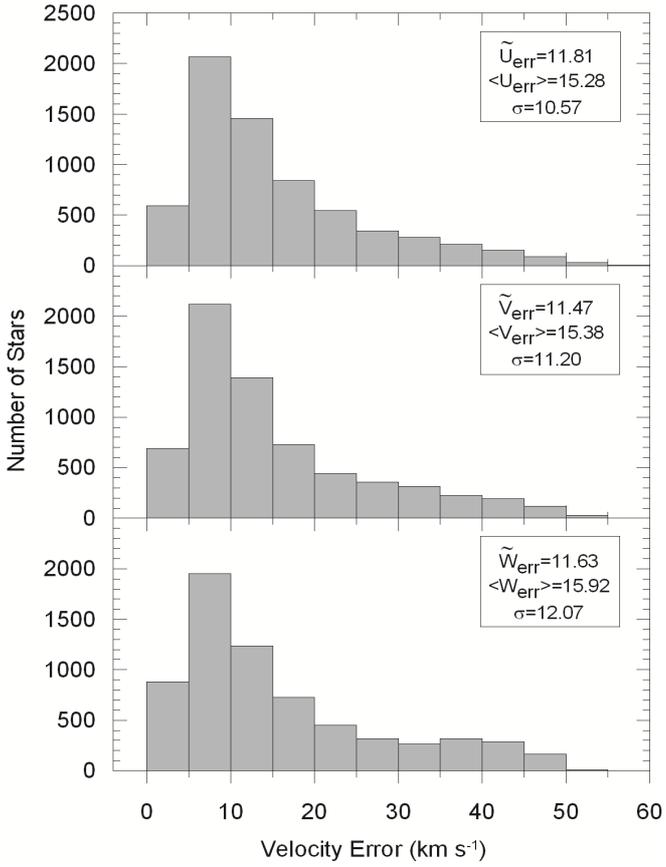}
\caption[] {Distribution of the errors of the space velocity components 
$U$ (upper panel), $V$ (middle panel) and $W$ (lower panel) for 
the sample used in this study. The medians, means and the standard 
deviations are also stated in the panels, respectively.}
\end{center}
\end{figure} 

\begin{figure}
\begin{center}
\includegraphics[scale=0.43, angle=0]{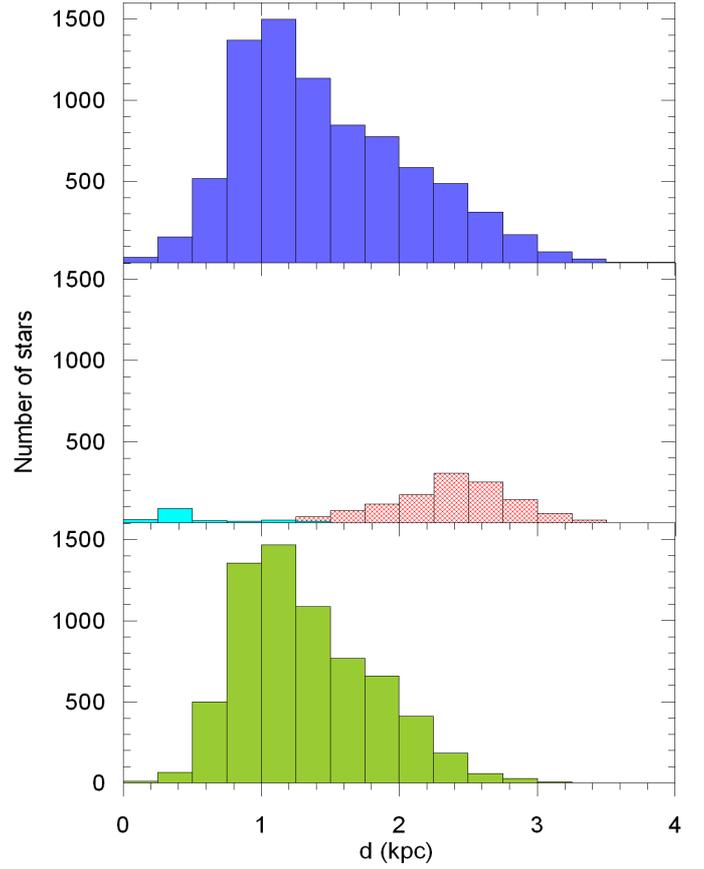}
\caption[] {Distance histogram of the RC stars. Upper panel corresponds to 
all RC stars ($N$=7985), while the lower one is drawn for the final sample 
($N$=6610). The middle panel corresponds to the RC stars with large proper 
motion errors ($\mu_{tot_{err}}>10$ mas yr$^{-1}$) and large distances 
($d>1$ kpc, right histogram, $N$=1204), and those close to the Galactic plane
($|b|\leq10^{o}$, left histogram, $N$=171).}
\end{center}
\end{figure} 

\begin{figure}
\begin{center}
\includegraphics[scale=0.43, angle=0]{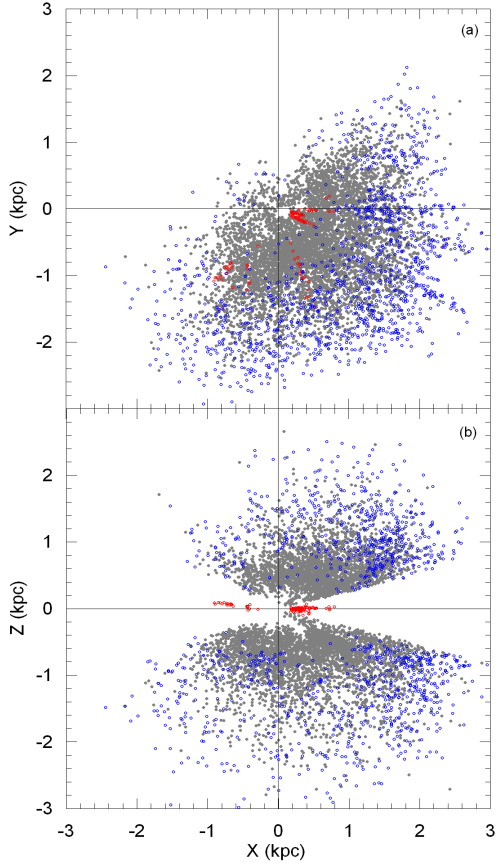}
\caption[] {Distribution of the sample stars (grey colour) and rejected ones: 
stars with $S_{err}>75$ kms$^{-1}$ (blue colour) and stars close to the 
Galactic plane (red colour) in the ($X$, $Y$) and ($X$, $Z$) planes. RC stars 
with errors $S_{err}>75$ kms$^{-1}$ dominate the outermost region of the diagram.}
\end{center}
\end{figure} 

\begin{figure}
\includegraphics[scale=0.43, angle=0]{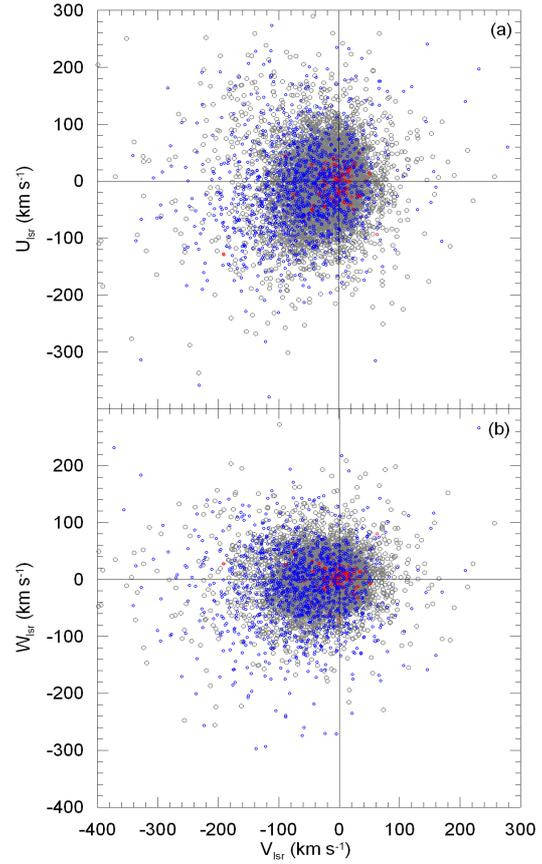}
\caption[] {Distribution of the sample stars (grey colour) and rejected ones: 
stars with $S_{err}>75$ kms$^{-1}$ (blue colour) and stars close to the 
Galactic plane (red colour) in two panels of space velocity components: 
(a) ($V_{lsr}$, $U_{lsr}$) and (b) ($V_{lsr}$, $W_{lsr}$).}
\end{figure} 

\begin{landscape}  
\begin{figure}
\begin{center}
\includegraphics[scale=1.15, angle=0]{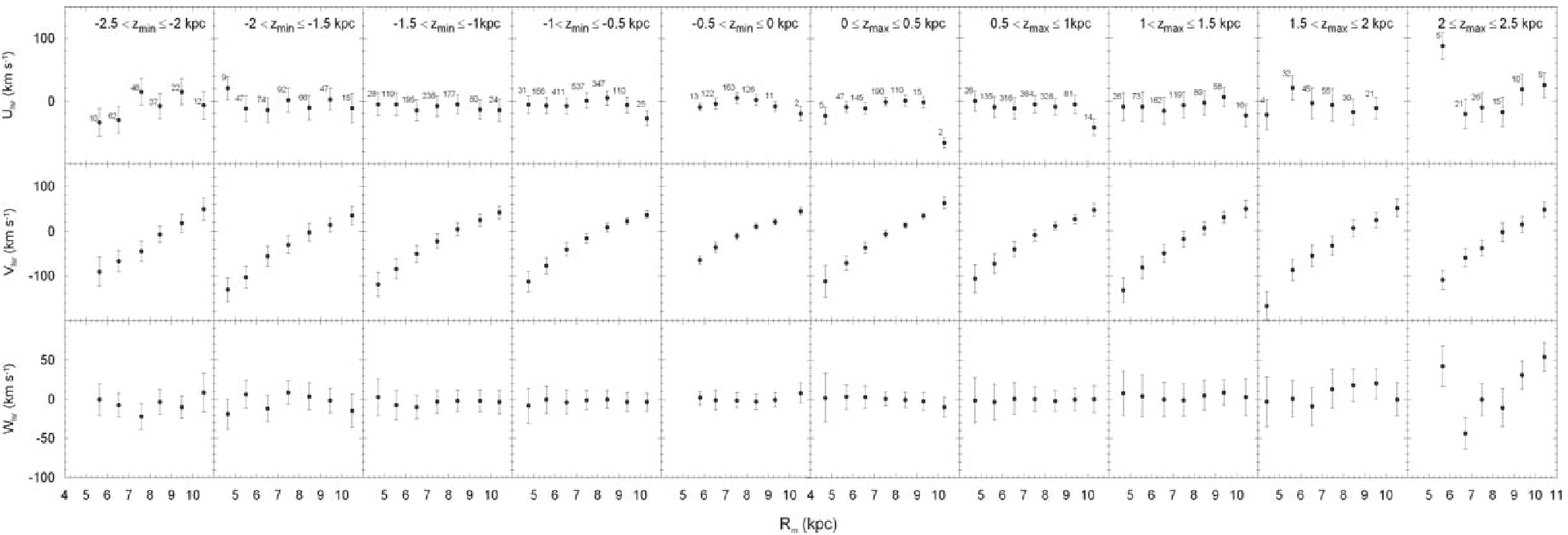}
\caption[] {Distribution of the space velocity components relative to the LSR 
velocities, $U_{lsr}$, $V_{lsr}$, $W_{lsr}$, in terms of mean Galactocentric 
radial distance $R_m$, for five $z_{min}$ and five $z_{max}$ intervals. The 
figures at the top panel indicate the number of stars. The error bars are 
also shown in the diagram.}
\end{center}
\end{figure} 
\end{landscape}

\begin{figure*}
\begin{center}
\includegraphics[scale=0.80, angle=0]{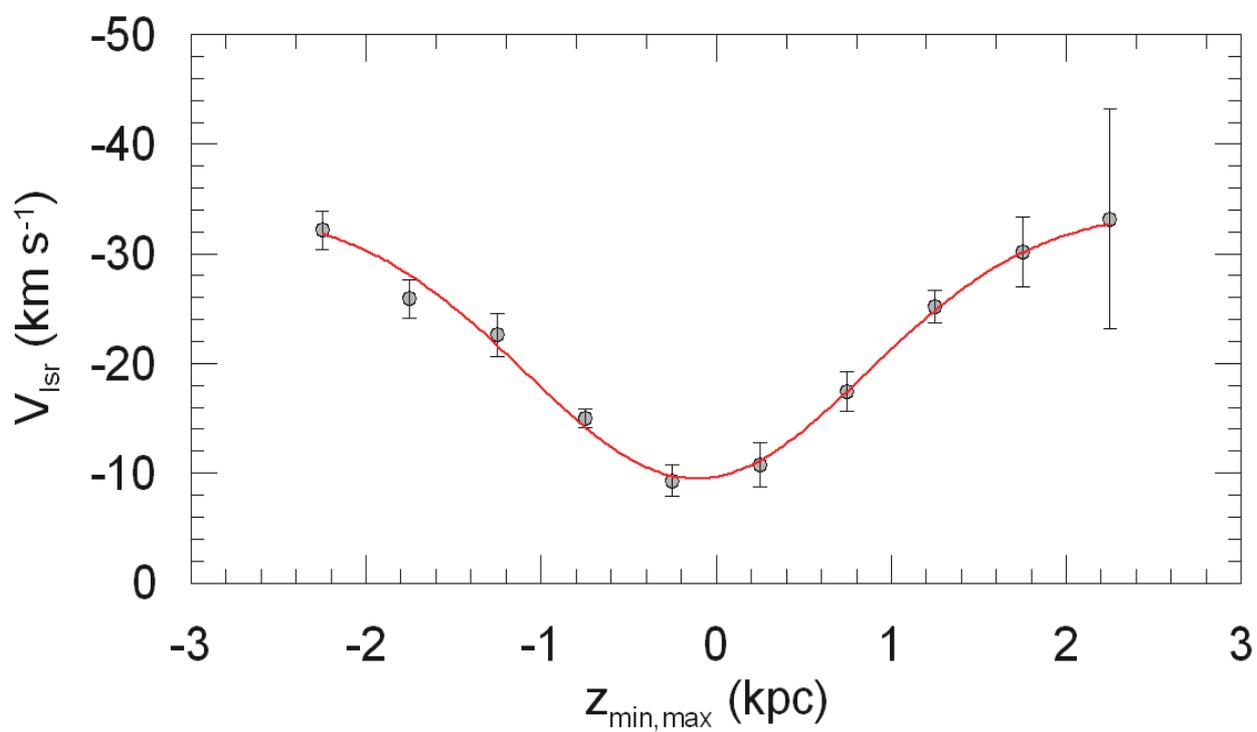}
\caption[] {Diagram of the mean $V_{lsr}$ space velocity components for five $z_{min}$ 
and five $z_{max}$ distances. The error bars are also shown in the diagram.}
\end{center}
\end{figure*} 

\begin{landscape}  
\begin{figure}
\begin{center}
\includegraphics[scale=1.15, angle=0]{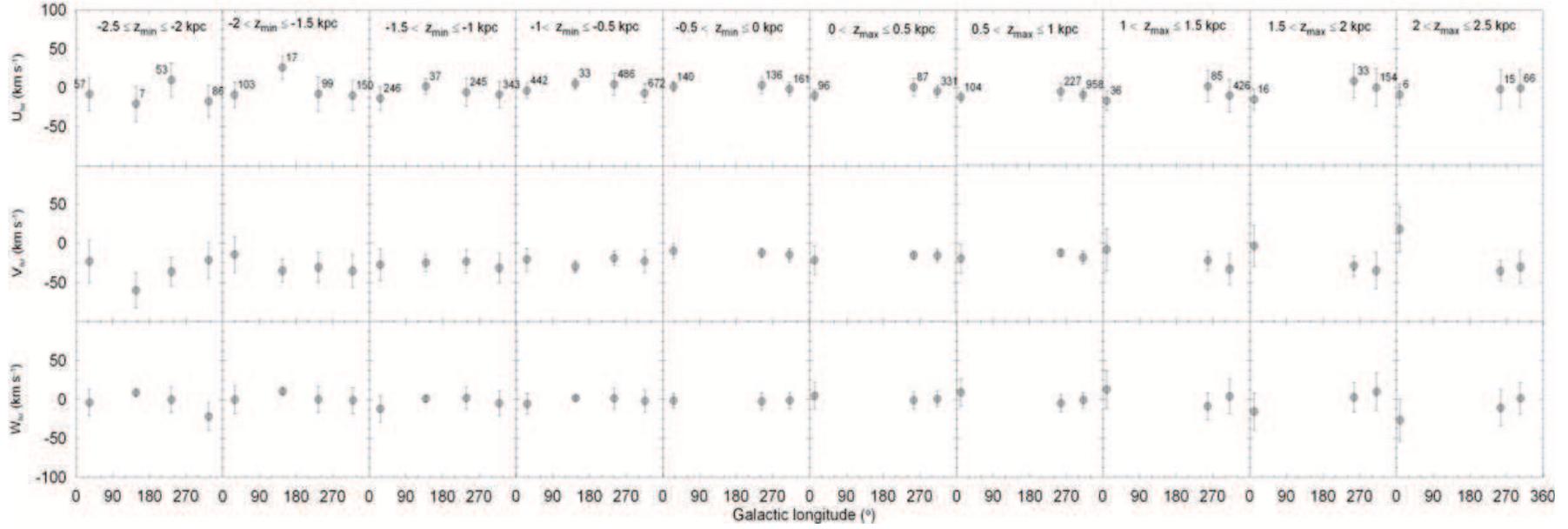}
\caption[] {Distribution of the space velocity components relative to the 
LSR velocities, $U_{lsr}$, $V_{lsr}$, $W_{lsr}$, in terms of Galactic 
longitude, for five $z_{min}$ and five $z_{max}$ intervals. The figures at 
the top panel indicate the number of stars. The error bars are also shown 
in the diagram.}
\end{center}
\end{figure}
\end{landscape} 

\begin{landscape}
\begin{figure}
\begin{center}
\includegraphics[scale=1.15, angle=0]{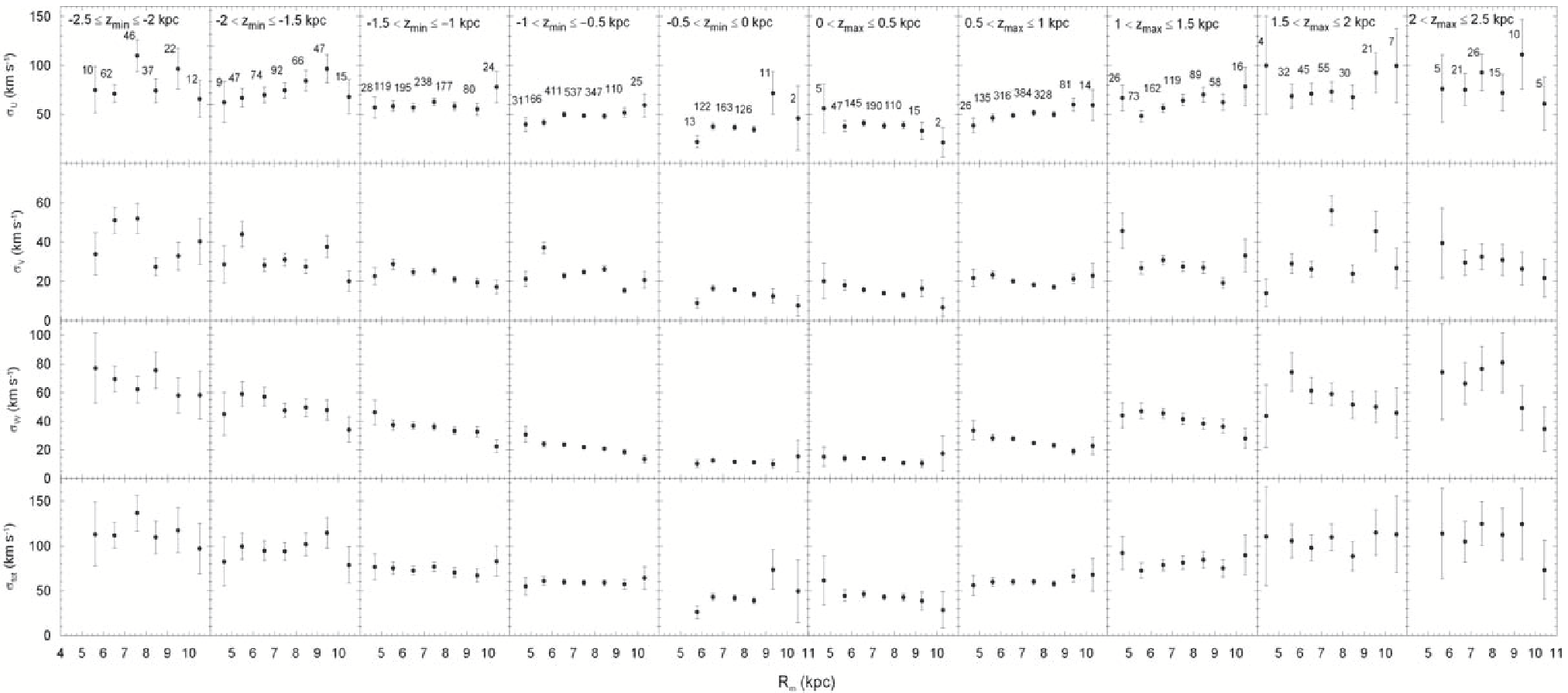}
\caption[] {Distribution of the dispersions of space velocity components 
relative to the LSR velocities, $\sigma_{U}$, $\sigma_{V}$, $\sigma_{W}$, 
$\sigma_{tot}$, in terms of mean Galactocentric radial distance $R_m$, 
for five $z_{min}$ and five $z_{max}$ intervals. The figures at the top 
panel indicate the number of stars. The error bars are also shown in the 
diagram.}
\end{center}
\end{figure} 
\end{landscape}

\begin{figure*}
\begin{center}
\includegraphics[scale=0.80, angle=0]{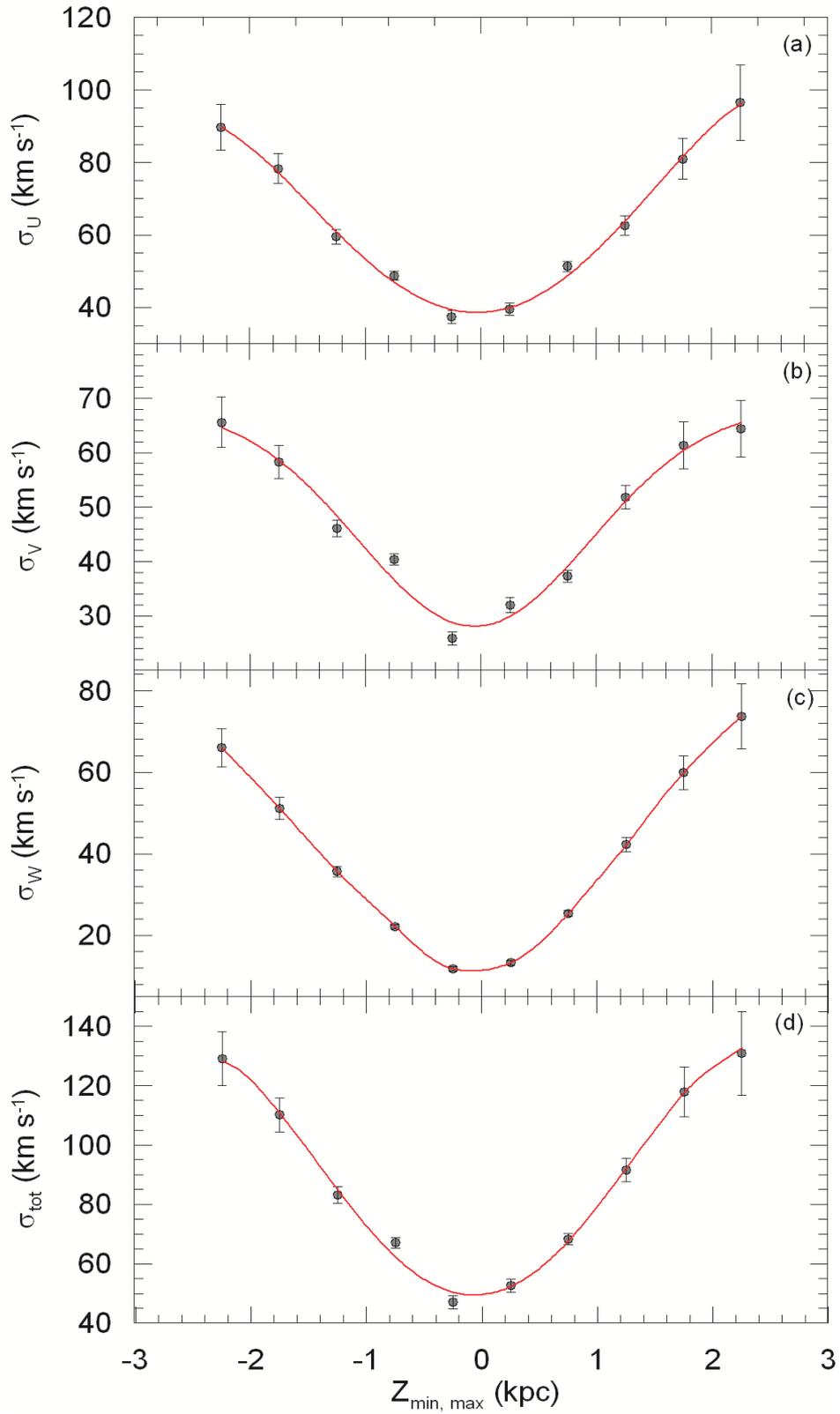}
\caption[] {Mean dispersions of the space velocity components, $\sigma_{U}$, 
$\sigma_{V}$, $\sigma_{W}$ and $\sigma_{tot}$, for five $z_{min}$ and 
five $z_{max}$ intervals.}
\end{center}
\end{figure*} 

\begin{figure*}
\begin{center}
\includegraphics[scale=.8, angle=0]{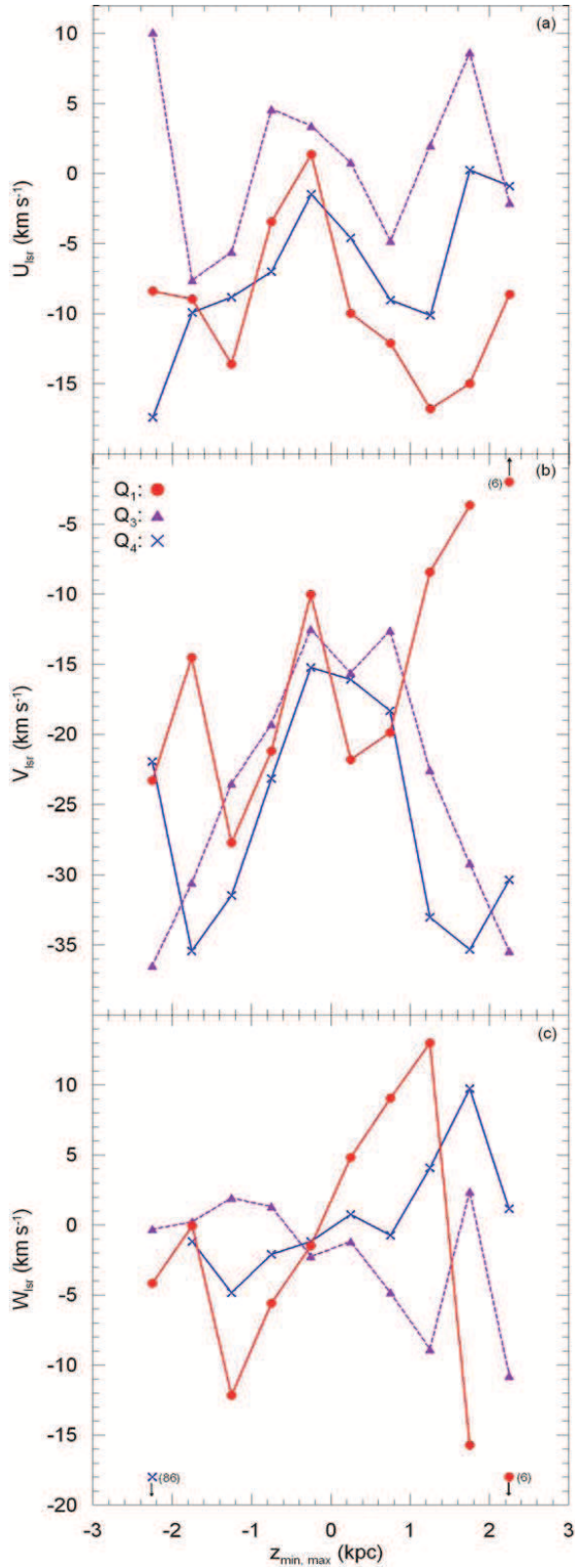}
\caption[] {Distribution of the space velocity components relative to the LSR 
velocities, $U_{lsr}$, $V_{lsr}$, $W_{lsr}$, for three quadrants: 
$Q_1(0^{o}\leq l\leq90^{o})$, $Q_3(180^{o}<l\leq270^{o})$ and $Q_4(270^{o}<l\leq360^{o})$ 
shown with different symbols, in terms of for five $z_{min}$ and five $z_{max}$ distances. 
Figures in the parenthesis indicate the (small) number of stars for which the 
errors are large.}  
\end{center}
\end{figure*}

\begin{table*}
\setlength{\tabcolsep}{4pt}
\center 
\scriptsize{
\caption{The space velocity components relative to the LSR velocities 
($U_{lsr}$, $V_{lsr}$, $W_{lsr}$) and their dispersions ($\sigma_U$, $\sigma_V$, 
$\sigma_W$, $\sigma_{tot}$), for the RC stars above ($b>10^{o}$) and below 
($b<-10^{o}$) the Galactic plane in terms of vertical eccentricity $e_v$. 
$N$ denotes the number of stars and, $\sigma_{tot}$ is the total space 
velocity dispersion. The errors of the space velocity components are the 
mean of the errors of the corresponding stars, while those for the dispersions 
are the standard errors.}
\begin{tabular}{lrrrrrrrrr}
\hline
 RC Sample & \multicolumn{1}{c}{$e_v$}  & \multicolumn{1}{c}{$N$} &    $<U_{lsr}>$ &   $<V_{lsr}>$ &    $<W_{lsr}>$ & \multicolumn{1}{c}{$\sigma_U$} & \multicolumn{1}{c}{$\sigma_V$} & \multicolumn{1}{c}{$\sigma_W$} & \multicolumn{1}{c}{$\sigma_{tot}$} \\
           &          &     & (kms$^{-1}$) & (kms$^{-1}$)& (kms$^{-1}$) & \multicolumn{1}{c}{(kms$^{-1}$)} & \multicolumn{1}{c}{(kms$^{-1}$)} & \multicolumn{1}{c}{(kms$^{-1}$)} & \multicolumn{1}{c}{(kms$^{-1}$)} \\
\hline
       All & [0, 0.12]    & 3283 & -3.16$\pm$11.58 &  -9.47$\pm$10.92 & -0.69$\pm$13.35 & 47.55$\pm$0.83 & 32.81$\pm$0.57 & 19.55$\pm$0.34 &  60.99$\pm$1.06 \\
           & (0.12, 0.25] & 2387 & -8.46$\pm$17.55 & -29.56$\pm$18.20 & -1.16$\pm$17.92 & 62.68$\pm$1.28 & 50.22$\pm$1.03 & 38.89$\pm$0.80 &  89.24$\pm$1.83 \\
           & (0.25, 1]    &  940 & -2.80$\pm$22.45 & -71.36$\pm$23.84 & -2.16$\pm$19.86 & 93.33$\pm$3.04 & 82.51$\pm$2.69 & 88.73$\pm$2.89 & 152.94$\pm$4.99 \\
$b>10^{o}$ & [0,0.12]     & 1550 & -6.65$\pm$11.99 &  -9.10$\pm$11.22 &  0.49$\pm$14.20 & 48.96$\pm$1.24 & 32.31$\pm$0.82 & 20.46$\pm$0.52 &  62.13$\pm$1.58 \\
           & (0.12, 0.25] &  929 & -7.53$\pm$20.23 & -32.71$\pm$19.75 &  2.15$\pm$22.21 & 63.33$\pm$2.08 & 52.56$\pm$1.72 & 41.42$\pm$1.36 &  92.14$\pm$3.02 \\
           & (0.25, 1]    &  330 & ~2.18$\pm$23.11 & -67.24$\pm$22.59 &  4.76$\pm$23.15 & 98.04$\pm$5.40 & 85.17$\pm$4.69 & 98.32$\pm$5.41 & 162.89$\pm$8.97 \\
$b<-10^{o}$& [0, 0.12]    & 1733 & -0.05$\pm$11.22 &  -9.80$\pm$10.65 & -1.74$\pm$12.58 & 46.06$\pm$1.11 & 33.25$\pm$0.80 & 18.63$\pm$0.45 &  59.78$\pm$1.44 \\
           & (0.12, 0.25] & 1458 & -9.05$\pm$15.85 & -27.56$\pm$17.20 & -3.26$\pm$15.19 & 62.27$\pm$1.63 & 48.59$\pm$1.27 & 37.04$\pm$0.97 &  87.24$\pm$2.28 \\
           & (0.25, 1]    &  610 & -5.50$\pm$22.10 & -73.59$\pm$24.52 & -5.90$\pm$18.08 & 90.64$\pm$3.67 & 81.03$\pm$3.28 & 82.94$\pm$3.36 & 147.18$\pm$5.96 \\
\hline
\end{tabular}
}  
\end{table*} 

\begin{table*}
\setlength{\tabcolsep}{3pt}
\center
\scriptsize{
\caption{The space velocity components relative to the LSR velocities 
($U_{lsr}$, $V_{lsr}$, $W_{lsr}$) and their space velocity dispersions 
($\sigma_U$, $\sigma_V$, $\sigma_W$, $\sigma_{tot}$) for the RC stars in terms 
of mean Galactocentric radial distance ($R_m$) for five $z_{min}$ and five $z_{max}$ 
intervals. $N$ denotes the number of RC stars in the $R_m$ range stated on the 
same line. Errors  as defined in Table 1.} 
\begin{tabular}{ccccrrrrrrr}
\hline
$z_{min}$/$z_{max}$& $R_m$ Range    &      $R_m$ &          $N$ & $<U_{lsr}>$ & $<V_{lsr}>$ & $<W_{lsr}>$ & \multicolumn{1}{c}{$\sigma_U$} & \multicolumn{1}{c}{$\sigma_V$} &  \multicolumn{1}{c}{$\sigma_W$}&  \multicolumn{1}{c}{$\sigma_{tot}$}\\
  (kpc)&   (kpc)& (kpc) &  & (kms$^{-1}$) & (kms$^{-1}$) & (kms$^{-1}$) & \multicolumn{1}{c}{(kms$^{-1}$)} & \multicolumn{1}{c}{(kms$^{-1}$)} &  \multicolumn{1}{c}{(kms$^{-1}$)} &  \multicolumn{1}{c}{(kms$^{-1}$)}\\
\hline
  (2, 2.5] &  (5, 6] &       5.65 &          5 & 87.75$\pm$21.54 & -109.10$\pm$21.14 & 42.48$\pm$25.49 & 76.37$\pm$34.15 & 39.45$\pm$17.64 & 74.49$\pm$33.31 & 113.74$\pm$50.87 \\
           &  (6, 7] &       6.72 &         21 & -20.70$\pm$23.23 & -59.97$\pm$20.72 & -43.58$\pm$20.21 & 75.35$\pm$16.44 & 29.47$\pm$6.43 & 66.43$\pm$14.50 & 104.69$\pm$22.85 \\
           &  (7, 8] &       7.49 &         26 & -10.28$\pm$22.87 & -38.39$\pm$18.05 & 0.04$\pm$20.08 & 93.01$\pm$18.24 & 32.58$\pm$6.39 & 76.79$\pm$15.06 & 124.94$\pm$24.50 \\
           &  (8, 9] &       8.47 &         15 & -17.38$\pm$23.26 & -2.93$\pm$20.96 & -10.69$\pm$24.03 & 72.31$\pm$18.67 & 30.82$\pm$7.96 & 80.87$\pm$20.88 & 112.78$\pm$29.12 \\
           & (9, 10] &       9.38 &         10 & 18.52$\pm$24.06 & 13.91$\pm$18.27 & 31.05$\pm$17.87 & 111.28$\pm$35.19 & 26.55$\pm$8.40 & 49.30$\pm$15.59 & 124.57$\pm$39.39 \\
           & 10, 11] &      10.42 &          5 & 25.33$\pm$19.93 & 47.42$\pm$17.32 & 54.09$\pm$17.97 & 61.00$\pm$27.28 & 21.66$\pm$9.69 & 34.49$\pm$15.42 & 73.35$\pm$32.80 \\
\hline
(1.5, 2]   &  (4, 5] &       4.40 &          4 & -21.85$\pm$24.20 & -167.26$\pm$31.43 & -3.21$\pm$31.57 & 100.24$\pm$50.12 & 14.10$\pm$7.05 & 43.76$\pm$21.88 & 110.28$\pm$55.14 \\
           &  (5, 6] &       5.61 &         32 & 21.28$\pm$19.59 & -87.49$\pm$23.86 & 0.82$\pm$23.30 & 69.03$\pm$12.20 & 28.99$\pm$5.12 & 74.33$\pm$13.14 & 105.50$\pm$18.65 \\
           &  (6, 7] &       6.53 &         45 & -3.31$\pm$24.62 & -55.30$\pm$23.98 & -9.15$\pm$24.30 & 71.31$\pm$10.63 & 26.23$\pm$3.91 & 61.53$\pm$9.17 & 97.77$\pm$14.57 \\
           &  (7, 8] &       7.48 &         55 & -6.42$\pm$26.03 & -32.82$\pm$20.83 & 13.27$\pm$24.86 & 73.20$\pm$9.87 & 56.06$\pm$7.56 & 59.11$\pm$7.97 & 109.52$\pm$14.77 \\
           &  (8, 9] &       8.45 &         30 & -17.49$\pm$20.53 & 6.28$\pm$19.52 & 18.22$\pm$20.73 & 67.86$\pm$12.39 & 23.99$\pm$4.38 & 51.57$\pm$9.42 & 88.54$\pm$16.17 \\
           & (9, 10] &       9.53 &         21 & -11.06$\pm$17.06 & 23.68$\pm$16.98 & 20.10$\pm$19.22 & 92.66$\pm$20.22 & 45.67$\pm$9.97 & 50.23$\pm$10.96 & 114.87$\pm$25.07 \\
           &(10, 11] &      10.51 &          7 & 50.08$\pm$22.16 & 50.89$\pm$19.59 & 0.13$\pm$21.06 & 99.80$\pm$37.72 & 26.72$\pm$10.10 & 45.85$\pm$17.33 & 113.03$\pm$42.72 \\
\hline
(1, 1.5]   &  (4, 5] &       4.68 &         26 & -9.09$\pm$21.74 & -132.42$\pm$27.69 & 7.97$\pm$28.03 & 66.87$\pm$13.11 & 45.85$\pm$8.99 & 44.11$\pm$8.65 & 92.30$\pm$18.10 \\
           &  (5, 6] &       5.57 &         73 & -9.22$\pm$23.60 & -81.78$\pm$25.03 & 4.13$\pm$26.89 & 48.12$\pm$5.63 & 26.82$\pm$3.14 & 47.25$\pm$5.53 & 72.58$\pm$8.49 \\
           &  (6, 7] &       6.58 &        162 & -15.33$\pm$21.32 & -49.40$\pm$20.22 & -0.01$\pm$22.09 & 56.33$\pm$4.43 & 30.86$\pm$2.42 & 45.53$\pm$3.58 & 78.73$\pm$6.19 \\
           &  (7, 8] &       7.50 &        119 & -6.78$\pm$20.45 & -18.52$\pm$17.79 & -0.92$\pm$20.68 & 64.63$\pm$5.92 & 27.44$\pm$2.52 & 41.59$\pm$3.81 & 81.61$\pm$7.48 \\
           &  (8, 9] &       8.47 &         89 & -2.66$\pm$19.02 & 6.11$\pm$15.15 & 5.11$\pm$19.26 & 70.38$\pm$7.46 &  27.00$\pm$2.86 & 38.29$\pm$4.06 & 84.55$\pm$8.96 \\
           & (9, 10] &       9.39 &         58 &  6.45$\pm$15.26 & 30.79$\pm$12.71 & 9.03$\pm$15.92 & 62.65$\pm$8.23 & 19.21$\pm$2.52 & 36.48$\pm$4.79 &  75.00$\pm$9.85 \\
           &(10, 11] &      10.42 &         16 & -23.04$\pm$17.50 & 49.08$\pm$19.57 & 2.91$\pm$23.18 & 78.52$\pm$19.63 & 33.13$\pm$8.28 & 28.13$\pm$7.03 & 89.75$\pm$22.44 \\
\hline
(0.5, 1]   &  (4, 5] &       4.71 &         26 & 0.09$\pm$16.16 & -106.82$\pm$31.25 & -1.11$\pm$28.50 & 38.71$\pm$7.59 & 21.82$\pm$4.28 & 33.71$\pm$6.61 & 55.78$\pm$10.94 \\
           &  (5, 6] &       5.62 &        135 & -9.55$\pm$16.21 & -73.10$\pm$21.83 & -3.33$\pm$22.89 & 46.61$\pm$4.01 & 23.35$\pm$2.01 & 28.39$\pm$2.44 & 59.36$\pm$5.11 \\
           &  (6, 7] &       6.56 &        316 & -11.64$\pm$16.77 & -40.25$\pm$16.85 & 0.73$\pm$20.41 &  49.00$\pm$2.76 & 20.15$\pm$1.13 & 27.79$\pm$1.56 & 59.83$\pm$3.37 \\
           &  (7, 8] &       7.53 &        384 & -4.99$\pm$13.72 & -9.72$\pm$12.88 & 0.23$\pm$15.64 & 51.52$\pm$2.63 & 18.28$\pm$0.93 & 24.74$\pm$1.26 &  60.00$\pm$3.06 \\
           &  (8, 9] &       8.47 &        328 & -9.10$\pm$13.16 & 11.08$\pm$8.84 & -2.04$\pm$12.84 & 49.78$\pm$2.75 & 17.06$\pm$0.94 & 23.28$\pm$1.29 & 57.54$\pm$3.18 \\
           & (9, 10] &       9.39 &         81 & -5.25$\pm$14.77 & 26.17$\pm$9.34 & -0.02$\pm$14.47 & 59.93$\pm$6.66 & 21.30$\pm$2.37 & 19.05$\pm$2.12 & 66.39$\pm$7.38 \\
           &(10, 11] &      10.30 &         14 & -41.99$\pm$13.13 & 47.01$\pm$13.93 & 0.34$\pm$16.98 & 59.35$\pm$15.86 & 22.98$\pm$6.14 & 22.85$\pm$6.11 & 67.62$\pm$18.07 \\
\hline
(0, 0.5]   &  (4, 5] &       4.70 &          5 & -22.77$\pm$13.34 & -112.67$\pm$35.09 & 1.83$\pm$30.94 & 56.06$\pm$25.07 & 20.29$\pm$9.07 & 15.16$\pm$6.78 & 61.52$\pm$27.51 \\
           &  (5, 6] &       5.67 &         47 & -9.27$\pm$8.57 & -71.55$\pm$15.84 & 3.16$\pm$15.42 & 38.11$\pm$5.56 & 18.19$\pm$2.65 & 14.11$\pm$2.06 & 44.52$\pm$6.49 \\
           &  (6, 7] &       6.56 &        145 & -11.24$\pm$9.38 & -37.70$\pm$12.84 & 2.92$\pm$14.55 &   41.00$\pm$3.40 & 15.79$\pm$1.31 & 14.36$\pm$1.19 & 46.22$\pm$3.84 \\
           &  (7, 8] &       7.52 &        190 &  -1.00$\pm$6.88 & -7.26$\pm$7.10 & 0.83$\pm$8.87 & 38.27$\pm$2.78 & 14.07$\pm$1.02 & 13.59$\pm$0.99 & 42.98$\pm$3.12 \\
           &  (8, 9] &       8.43 &        110 &   1.04$\pm$9.34 & 12.48$\pm$5.57 & -0.57$\pm$9.79 & 39.01$\pm$3.72 & 13.08$\pm$1.25 & 10.76$\pm$1.03 & 42.53$\pm$4.06 \\
           & (9, 10] &       9.29 &         15 & -1.90$\pm$9.15 & 33.36$\pm$7.08 & -2.68$\pm$11.32 & 33.19$\pm$8.57 & 16.41$\pm$4.24 & 10.49$\pm$2.71 & 38.48$\pm$9.94 \\
           &(10, 11] &      10.28 &          2 & -65.76$\pm$7.95 & 62.4$\pm$13.49 & -9.95$\pm$13.02 & 21.50$\pm$15.20 & 6.87$\pm$4.86 & 17.49$\pm$12.37 & 28.55$\pm$20.19 \\
\hline
(-0.5, 0]  &  (5, 6] &       5.79 &         13 & -8.91$\pm$6.07 & -65.01$\pm$9.26 & 1.73$\pm$8.76 & 22.23$\pm$6.17 & 8.95$\pm$2.48 & 10.36$\pm$2.87 & 26.11$\pm$7.24 \\
           &  (6, 7] &       6.53 &        122 & -3.97$\pm$9.11 & -36.73$\pm$11.07 & -1.15$\pm$12.63 & 37.68$\pm$3.41 & 16.47$\pm$1.49 & 12.58$\pm$1.14 &  43.00$\pm$3.89 \\
           &  (7, 8] &       7.53 &        163 & 5.24$\pm$8.42 & -11.46$\pm$7.24 & -1.22$\pm$9.31 & 36.78$\pm$2.88 & 15.78$\pm$1.24 & 11.65$\pm$0.91 & 41.68$\pm$3.26 \\
           &  (8, 9] &       8.44 &        126 & 2.31$\pm$8.53 & 10.52$\pm$6.91 & -3.07$\pm$9.76 & 34.66$\pm$3.09 & 13.36$\pm$1.19 & 11.39$\pm$1.01 & 38.85$\pm$3.46 \\
           & (9, 10] &       9.33 &         11 & -8.04$\pm$7.28 & 20.60$\pm$6.71 & -0.80$\pm$9.21 & 72.04$\pm$21.72 & 12.56$\pm$3.79 & 10.13$\pm$3.05 & 73.83$\pm$22.26 \\
           &(10, 11] &      10.52 &          2 & -19.57$\pm$11.25 & 44.18$\pm$8.22 & 7.98$\pm$12.53 & 46.29$\pm$32.73 & 7.57$\pm$5.35 & 15.61$\pm$11.04 & 49.43$\pm$34.95 \\
\hline
(-1, -0.5] &  (4, 5] &       4.75 &         31 & -5.68$\pm$13.99 & -113.06$\pm$22.80 & -8.44$\pm$22.34 & 39.79$\pm$7.15 & 21.27$\pm$3.82 & 30.92$\pm$5.55 & 54.70$\pm$9.82 \\
           &  (5, 6] &       5.60 &        166 & -7.04$\pm$12.61 & -77.24$\pm$18.00 & -0.40$\pm$17.08 & 41.71$\pm$3.24 & 37.16$\pm$2.88 & 24.04$\pm$1.87 & 60.82$\pm$4.72 \\
           &  (6, 7] &       6.55 &        411 & -7.90$\pm$12.65 & -40.74$\pm$15.02 & -3.76$\pm$15.35 & 49.91$\pm$2.46 & 22.87$\pm$1.13 & 23.55$\pm$1.16 & 59.74$\pm$2.95 \\
           &  (7, 8] &       7.48 &        537 & 1.06$\pm$11.86 & -15.97$\pm$10.99 & -1.09$\pm$12.41 & 48.50$\pm$2.09 & 24.81$\pm$1.07 & 21.89$\pm$0.94 & 58.71$\pm$2.53 \\
           &  (8, 9] &       8.44 &        347 & 5.02$\pm$11.48 & 8.06$\pm$9.55 & -0.38$\pm$11.44 & 48.22$\pm$2.59 & 26.35$\pm$1.41 & 20.73$\pm$1.11 & 58.73$\pm$3.15 \\
           & (9, 10] &       9.38 &        110 & -6.06$\pm$12.49 & 22.12$\pm$8.41 & -3.48$\pm$12.00 & 51.75$\pm$4.93 & 15.52$\pm$1.48 & 18.48$\pm$1.76 & 57.10$\pm$5.44 \\
           &(10, 11] &      10.31 &         25 & -27.19$\pm$12.18 & 36.35$\pm$8.49 & -3.45$\pm$11.36 & 59.25$\pm$11.85 & 20.73$\pm$4.15 & 13.41$\pm$2.68 & 64.19$\pm$12.84 \\
\hline
(-1.5, -1] &  (4, 5] &       4.69 &         28 & -4.78$\pm$17.73 & -119.04$\pm$26.82 & 2.72$\pm$22.81 & 56.99$\pm$10.77 & 22.66$\pm$4.28 & 46.11$\pm$8.71 & 76.73$\pm$14.5 \\
           &  (5, 6] &       5.55 &        119 & -4.98$\pm$18.10 & -84.04$\pm$22.36 & -7.68$\pm$18.61 & 58.48$\pm$5.36 & 28.75$\pm$2.64 & 37.41$\pm$3.43 & 75.14$\pm$6.89 \\
           &  (6, 7] &       6.50 &        195 & -14.39$\pm$16.40 & -50.73$\pm$18.03 & -10.22$\pm$15.20 & 57.17$\pm$4.09 & 24.81$\pm$1.78 & 36.95$\pm$2.65 & 72.45$\pm$5.19 \\
           &  (7, 8] &       7.47 &        238 & -7.80$\pm$16.24 & -22.32$\pm$16.47 & -3.22$\pm$14.16 & 62.92$\pm$4.08 & 25.41$\pm$1.65 & 36.15$\pm$2.34 & 76.89$\pm$4.98 \\
           &  (8, 9] &       8.43 &        177 & -5.13$\pm$15.42 & 4.04$\pm$14.28 & -1.95$\pm$13.61 & 58.27$\pm$4.38 &  21.00$\pm$1.58 & 33.4$\pm$2.51 & 70.37$\pm$5.29 \\
           & (9, 10] &       9.48 &         80 & -13.27$\pm$15.01 & 24.26$\pm$14.22 & -1.84$\pm$13.98 & 55.36$\pm$6.19 & 19.37$\pm$2.17 & 32.57$\pm$3.64 & 67.09$\pm$7.50 \\
           &(10, 11] &      10.39 &         24 & -14.27$\pm$18.00 & 41.17$\pm$14.35 & -3.65$\pm$15.00 & 78.03$\pm$15.93 & 17.22$\pm$3.52 & 22.56$\pm$4.61 & 83.03$\pm$16.95 \\
\hline
(-2, -1.5] &  (4, 5] &       4.64 &          9 & 21.07$\pm$18.58 & -130.50$\pm$26.48 & -18.96$\pm$19.23 & 62.66$\pm$20.89 & 28.67$\pm$9.56 & 45.14$\pm$15.05 & 82.38$\pm$27.46 \\
           &  (5, 6] &       5.49 &         47 & -11.93$\pm$20.65 & -103.02$\pm$24.63 & 6.07$\pm$17.65 & 66.94$\pm$9.76 & 44.11$\pm$6.43 & 58.99$\pm$8.60 & 99.53$\pm$14.52 \\
           &  (6, 7] &       6.52 &         74 & -14.17$\pm$19.40 & -55.66$\pm$22.21 & -11.54$\pm$16.94 & 69.89$\pm$8.12 & 28.29$\pm$3.29 & 57.20$\pm$6.65 & 94.64$\pm$11.00 \\
           &  (7, 8] &       7.48 &         92 & 2.11$\pm$19.06 & -30.44$\pm$19.63 & 8.79$\pm$14.73 & 74.77$\pm$7.80 & 31.08$\pm$3.24 & 47.60$\pm$4.96 & 93.93$\pm$9.79 \\
           &  (8, 9] &       8.46 &         66 & -10.47$\pm$19.54 & -2.70$\pm$20.14 & 3.86$\pm$17.04 & 84.45$\pm$10.40 & 27.53$\pm$3.39 & 49.68$\pm$6.12 & 101.77$\pm$12.53 \\
           & (9, 10] &       9.46 &         47 &   3.14$\pm$17.15 & 13.82$\pm$15.68 & -1.56$\pm$15.42 & 96.67$\pm$14.10 & 37.73$\pm$5.50 & 47.95$\pm$6.99 & 114.31$\pm$16.67 \\
           &(10, 11] &      10.48 &         15 & -10.94$\pm$23.36 & 34.35$\pm$20.91 & -14.55$\pm$21.41 & 68.15$\pm$17.60 & 20.17$\pm$5.21 & 34.15$\pm$8.82 & 78.85$\pm$20.36 \\
\hline
(-2.5, -2] &  (5, 6] &       5.62 &         10 & -33.86$\pm$22.22 & -90.71$\pm$32.38 & -0.23$\pm$19.88 & 75.15$\pm$23.76 & 34.02$\pm$10.76 & 77.27$\pm$24.43 & 113.03$\pm$35.74 \\
           &  (6, 7] &       6.52 &         62 & -29.86$\pm$21.52 & -66.94$\pm$23.89 & -7.64$\pm$15.43 & 71.38$\pm$9.07 & 51.02$\pm$6.48 & 69.59$\pm$8.84 & 111.99$\pm$14.22 \\
           &  (7, 8] &       7.59 &         46 & 15.04$\pm$21.28 & -45.19$\pm$22.63 & -22.06$\pm$16.38 & 109.82$\pm$16.19 & 52.06$\pm$7.68 & 62.26$\pm$9.18 & 136.55$\pm$20.13 \\
           &  (8, 9] &       8.46 &         37 & -7.69$\pm$20.36 & -7.12$\pm$18.28 & -3.64$\pm$15.91 & 74.40$\pm$12.23 & 27.42$\pm$4.51 & 75.53$\pm$12.42 & 109.51$\pm$18.00 \\
           & (9, 10] &       9.49 &         22 &  14.94$\pm$20.52 & 17.46$\pm$20.26 & -10.14$\pm$14.29 & 96.70$\pm$20.62 & 32.87$\pm$7.01 & 58.14$\pm$12.40 & 117.52$\pm$25.06 \\
           &(10, 11] &      10.51 &         12 & -6.83$\pm$22.10 & 49.11$\pm$25.90 & 8.53$\pm$24.78 & 65.98$\pm$19.05 & 40.3$\pm$11.63 & 58.33$\pm$16.84 & 96.85$\pm$27.96 \\
\hline
\end{tabular}  
}
\end{table*} 

\begin{table*}
\setlength{\tabcolsep}{3pt}
\center
\scriptsize{
\caption{The mean space velocity components relative to the LSR velocities 
($U_{lsr}$, $V_{lsr}$, $W_{lsr}$) for four quadrants and for five $z_{min}$ and 
five $z_{min}$ intervals. $N$ and $<l>$ denote the number of stars and the mean 
longitude for the corresponding quadrant. Errors as defined in Table 1.} 
\begin{tabular}{cccccccccccc}
\hline
 \multicolumn{6}{c}{$b>10^{o}$} &  \multicolumn{6}{c}{$b<-10^{o}$}\\
\hline
$z_{max}$  & $N$ &  $<l>$ & $<U_{lsr}>$ &  $<V_{lsr}>$ &  $<W_{lsr}>$ & $z_{min}$  &     $N$ &      $<l>$ &     $<U_{lsr}>$ &   $<V_{lsr}>$ &     $<W_{lsr}>$ \\
(kpc)&   &  ($^{o}$) & (kms$^{-1}$) &  (kms$^{-1}$) &  (kms$^{-1}$) & (kpc) & & ($^{o}$) & (kms$^{-1}$) & (kms$^{-1}$) & (kms$^{-1}$) \\
\hline
  [0, 0.5] &         96 &         12 & -9.99$\pm$7.29 & -21.80$\pm$18.04 & 4.82$\pm$17.44 &  (-0.5, 0] &        140 &         26 & 1.36$\pm$6.10 & -10.01$\pm$9.51 & -1.47$\pm$9.28 \\
           &        --- &        --- &        --- &        --- &        --- &            &        --- &        --- &        --- &        --- &        --- \\
           &         87 &        255 & 0.78$\pm$11.58 & -15.62$\pm$5.09 & -1.18$\pm$11.09 &            &        136 &        243 & 3.41$\pm$10.77 & -12.51$\pm$6.56 & -2.24$\pm$11.39 \\
           &        331 &        313 & -4.60$\pm$7.89 & -16.08$\pm$8.17 & 0.77$\pm$9.99 &            &        161 &        311 & -1.47$\pm$8.83 & -15.23$\pm$8.63 & -1.16$\pm$10.44 \\
  (0.5, 1] &        104 &         11 & -12.11$\pm$10.25 & -19.87$\pm$22.92 & 9.04$\pm$21.38 & (-1, -0.5] &        442 &         27 & -3.45$\pm$9.86 & -21.20$\pm$14.43 & -5.57$\pm$13.16 \\
           &        --- &        --- &        --- &        --- &        --- &            &         33 &        146 & 5.45$\pm$7.14 & -30.06$\pm$7.56 & 1.68$\pm$2.47 \\
           &        227 &        256 & -4.81$\pm$15.57 & -12.60$\pm$7.64 & -4.82$\pm$14.32 &            &        486 &        241 & 4.60$\pm$14.31 & -19.28$\pm$9.36 & 1.32$\pm$13.40 \\
           &        958 &        311 & -9.02$\pm$14.97 & -18.32$\pm$14.41 & -0.73$\pm$17.22 &            &        672 &        316 & -6.99$\pm$12.36 & -23.17$\pm$13.59 & -2.07$\pm$14.55 \\
  (1, 1.5] &         36 &          9 & -16.79$\pm$12.84 & -8.42$\pm$26.54 & 13.01$\pm$24.26 & (-1.5, -1] &        246 &         27 & -13.62$\pm$14.98 & -27.70$\pm$20.08 & -12.16$\pm$16.70 \\
           &        --- &        --- &        --- &        --- &        --- &            &         37 &        139 & 1.87$\pm$10.17 & -25.33$\pm$10.60 & 0.96$\pm$3.70 \\
           &         85 &        257 & 2.00$\pm$20.00 & -22.55$\pm$12.06 & -8.85$\pm$17.16 &            &        245 &        238 & -5.59$\pm$17.72 & -23.52$\pm$13.72 & 1.93$\pm$14.14 \\
           &        426 &        311 & -10.11$\pm$20.96 & -33.04$\pm$19.87 & 4.09$\pm$22.32 &            &        343 &        319 & -8.82$\pm$16.99 & -31.49$\pm$18.57 & -4.83$\pm$16.30 \\
  (1.5, 2] &         16 &         11 & -15.00$\pm$13.03 & -3.66$\pm$26.64 & -15.70$\pm$23.56 & (-2, -1.5] &        103 &         29 & -8.97$\pm$16.25 & -14.54$\pm$22.62 & -0.07$\pm$17.8 \\
           &        --- &        --- &        --- &        --- &        --- &            &         17 &        147 & 26.06$\pm$14.13 & -34.90$\pm$14.26 & 10.60$\pm$4.18 \\
           &         33 &        256 & 8.65$\pm$21.54 & -29.16$\pm$12.45 & 2.40$\pm$19.09 &            &         99 &        235 & -7.61$\pm$22.09 & -30.57$\pm$19.00 & 0.23$\pm$17.23 \\
           &        154 &        311 & 0.25$\pm$23.87 & -35.34$\pm$22.93 & 9.74$\pm$24.18 &            &        150 &        319 & -9.91$\pm$19.69 & -35.45$\pm$20.94 & -1.18$\pm$16.73 \\
  (2, 2.5] &          6 &          8 & -8.62$\pm$13.59 & 17.86$\pm$29.23 & -26.61$\pm$26.71 & (-2.5, -2] &         57 &         33 & -8.41$\pm$21.23 & -23.32$\pm$27.12 & -4.17$\pm$16.67 \\
           &        --- &        --- &        --- &        --- &        --- &            &          7 &        147 & -20.63$\pm$22.84 & -59.88$\pm$22.71 & 8.78$\pm$4.62 \\
           &         15 &        255 & -2.08$\pm$25.21 & -35.44$\pm$12.57 & -10.75$\pm$23.25 &            &         53 &        234 & 10.09$\pm$21.84 & -36.48$\pm$17.92 & -0.30$\pm$16.73 \\
           &         66 &        304 & -0.90$\pm$23.76 & -30.37$\pm$19.95 & 1.16$\pm$20.37 &            &         86 &        326 & -17.40$\pm$20.60 & -21.94$\pm$22.78 & -21.95$\pm$17.91 \\
\hline
\end{tabular}  
}
\end{table*} 

\end{document}